\newif\ifMNRAS
\MNRASfalse 

\PassOptionsToPackage{pdfpagelabels=false}{hyperref} 

\ifMNRAS
    \documentclass[fleqn,usenatbib,useAMS]{mnras}
    \usepackage{graphicx}
\else

\documentclass[]{aastex63}
\fi

\usepackage{acronym}
\usepackage{hyperref}
\usepackage{amsmath,amsfonts,amssymb}
\usepackage{mathrsfs}
\usepackage{mathtools}
\usepackage[utf8]{inputenc}

\usepackage[normalem]{ulem}

\usepackage{pifont}

\newcommand{\adam}[1]{{#1}}

\date{Accepted September 19, 2023. Received August 9, 2023}

\shorttitle{BH CCSNe}

\begin{document}
\title{Black-Hole Formation Accompanied by the Supernova Explosion of a 40-M$_{\odot}$ Progenitor Star}
\correspondingauthor{Adam Burrows}
\email{burrows@astro.princeton.edu}
\author[0000-0002-3099-5024]{Adam Burrows}
\affiliation{Department of Astrophysical Sciences, Princeton University, NJ 08544, USA; School of Natural Sciences, Institute for Advanced Study, Princeton, NJ 08540}
\author[0000-0003-1938-9282]{David Vartanyan}
\affiliation{Carnegie Observatories, 813 Santa Barbara St, Pasadena, CA 91101, USA; NASA Hubble Fellow}
\author[0000-0002-0042-9873]{Tianshu Wang}
\affiliation{Department of Astrophysical Sciences, Princeton University, NJ 08544, USA}

\begin{abstract}
We have simulated the collapse and evolution of the core of a solar-metallicity 40-M$_{\odot}$ star and find that it explodes vigorously by the neutrino mechanism. This despite its very high ``compactness". Within $\sim$1.5 seconds of explosion, a black hole forms. The explosion is very asymmetrical and has a total explosion energy of $\sim$1.6$\times$10$^{51}$ ergs. At black hole formation, its baryon mass is $\sim$2.434 M$_{\odot}$ and gravitational mass is 2.286 M$_{\odot}$. Seven seconds after black hole formation an additional $\sim$0.2 M$_{\odot}$ is accreted, leaving a black hole baryon mass of $\sim$2.63 M$_{\odot}$. A disk forms around the proto-neutron star, from which a pair of neutrino-driven jets emanates. These jets accelerate some of the matter up to speeds of $\sim$45,000 km s$^{-1}$ and contain matter with entropies of $\sim$50. The large spatial asymmetry in the explosion results in a residual black hole recoil speed of $\sim$1000 km s$^{-1}$. This novel black-hole formation channel now joins the other black-hole formation channel between $\sim$12 and $\sim$15 M$_{\odot}$ discovered previously and implies that the black-hole/neutron-star birth ratio for solar-metallicity stars could be $\sim$20\%. However, one channel leaves black holes in perhaps the $\sim$5-15 M$_{\odot}$ range with low kick speeds, while the other leaves black holes in perhaps the $\sim$2.5-3.0 M$_{\odot}$ mass range with high kick speeds. \adam{However, even $\sim$8.8 seconds after core bounce the newly-formed black hole is still accreting at a rate of $\sim$2$\times$10$^{-2}$ M$_{\odot}$ s$^{-1}$ and whether the black hole eventually achieves a significantly larger mass over time is yet to be determined.}

\end{abstract} 

\ifMNRAS
    \begin{keywords}
    stars - supernovae - general
    \end{keywords}
\else
    \keywords{
    stars - supernovae - general }
\fi

\section{Introduction}
\label{sec:int}

The terminal stages of the evolution of massive stars (ZAMS mass $\ge$ 8 M$_{\odot}$) end in the dynamical collapse
of their cores.  The residue of collapse is either a neutron star or a stellar-mass black hole, each central players in their own right in a variety of important astronomical contexts. As such, their origin and statistics are of abiding interest throughout astronomy \citep{faucher_kaspi,2023arXiv230409368M}. However, the mapping between initial progenitor star and final compact object has, despite decades of speculation and effort, remained only dimly perceived. Core-collapse and supernova theory \citep{1985ApJ...295...14B,burrows:95,janka2012,burrows2013,burrows_2020} addresses the context of their birth, but given that theory's complexity there have emerged a number of simple prescriptions and ``cookbooks," loosely informed by a general knowledge of the progenitor structures and supernova modeling, that purport to provide such a mapping \citep{zhang2008,2012ApJ...757...69U,2016ApJ...818..124E,2016MNRAS.460..742M,swbj16,2019ApJ...870....1E}. \citet{2011ApJ...730...70O} attempted a slightly more comprehensive attack on the mapping between progenitor mass and residue, using their 1D general-relativistic code GR1D and exploring the effect of nuclear equation of state. However, they too, due to the absence of the crucial multi-D effects, fell short of a credible predictive prescription. This multitude of prescriptions has jumped in to fill the gap that detailed state-of-the-art 3D simulations have until recently only slowly started to fill. 

However, this recent significant progress in the theory of core-collapse supernovae (CCSNe) \citep{burrows_2019,vartanyan2019,muller_lowmass,stockinger2020,burrows_2020,bollig2021,sandoval2021,nakamura2022,vartanyan2023} is calling into question the accuracy and usefulness of these simple prescriptions. Many invoke ``compactness" \citep{2011ApJ...730...70O} as an explodability condition,
suggesting that a low compactness (steep progenitor core mass density profile) is necessary for explosion by the neutrino mechanism \footnote{The compactness parameter very loosely characterizes the core structure and is defined as
\begin{equation}
\xi_M= \frac{M/M_{\odot}}{R(M)/1000\, \mathrm{km}}\,,
\end{equation}
where the subscript $M$ denotes the interior mass coordinate at which the compactness parameter is evaluated. For our purposes, we evaluate the compactness parameter $\xi_{1.75}$ at $M$ = 1.75 M$_{\odot}$.}. Recent sophisticated 3D CCSN simulations strongly indicate that this is false. Both low and high compactness models explode \citep{burrows_2020,bollig2021}.  Indeed, the neutrino mechanism requires higher compactness to obtain the canonical explosion energy of $\sim$10$^{51}$ ergs ($\equiv$ one Bethe) \citep{wang}. This is because the integrated neutrino heating upon the delayed launch of the explosion is roughly proportional to the neutrino luminosity times the ``optical depth" of the mantle to neutrino absorption, both of which are higher for higher compactness (linked to mass accretion rate).  Generally, higher-compactness models explode later after bounce, but more energetically. A slightly more sophisticated cookbook invokes the ``Ertl" prescription that calibrates the explosion energy to the Crab and SN1987A \citep{2016ApJ...818..124E,swbj16}. However, this too has been shown to be imperfect \citep{muller2016,burrows_2020,2021Natur.589...29B,wang,tsang2022}.  

Interestingly, these recipes have for the most part (with the exception of \citealt{2020ApJ...890..127C}) also failed to predict the formation of black holes at intermediate compactness that is seen in state-of-the-art 3D simulations \citep{burrows_2019,burrows_2020,vartanyan2023}, and also witnessed in 2D CCSN simulations \citep{tsang2022,wang} (see Figure \ref{fig:compactness-he}). In the ZAMS progenitor mass range of $\sim$12 to $\sim$15 M$_{\odot}$ and occupying an island or branch in the helium-core-mass/compactness distributions \citep{wang} from the \citet{swbj16} and \citet{sukhbold2018} massive star compendium, it is found that many models with weak silicon/oxygen interfaces and intermediate compactness fail to explode. If true, this would be the major black-hole formation channel. However, this conclusion is contingent upon the accuracy of both the CCSN theory that implies it and the massive star models that establish the overall initial structural context of core collapse at stellar death. There is still significant room for improvement in the theory of CCSN explosions. In addition, the end stage of massive star evolution manifests complicated burning shell mergers, overshoot, and mixing that are only now being studied in multi-D \citep{Chatzopoulos2016,muller2016,muller2017,Muller2019,takahashi2019,fields2020,Fields2021,2021MNRAS.506L..20Y,2021ApJ...908...44Y}. Nevertheless, this coupling of sophisticated progenitor models with state-of-the-art supernova theory is now clearly at odds with the bulk of the simple prescriptions used over a decade to determine the outcomes of collapse. 

Nowhere, perhaps, is this mismatch more apparent than in the notion embedded in these simple prescriptions that the highest compactness models can not explode. Related to this emerges the suggestion that there is a mass cut in ZAMS mass, roughly paralleling compactness at masses above $\sim$20 M$_{\odot}$, above which black holes form without 
experiencing a supernova explosion \citep{smartt2009,2009MNRAS.395.1409S} \footnote{The concept of  ``failed supernovae"  or ``unnovae" \citep{2008ApJ...684.1336K} is now lodged, however loosely, in astronomical lore. There is little doubt that some massive stars end their lives more quietly than those that lead to standard core-collapse supernovae \citep{gerke2015,adams2017a,adams2017b}. However, which stars these are has not been definitively determined by reliable theoretical means.}.   

With this paper, we challenge this notion. Using our CCSN code F{\sc{ornax}}, we have simulated in 3D the dynamical evolution of the core of the 40-M$_{\odot}$ solar-metallicity model of \citet{swbj16} \citep{wh07}. We find that this model explodes robustly $\sim$0.25 seconds after bounce and launches a vigorous and quite asymmetrical explosion. Due to simultaneous and continuing accretion and near $\sim$1.76 seconds after bounce, when the shock wave  achieves a radius near $\sim$20,000 kilometers (km) its proto-neutron star (PNS) core becomes unstable to the general-relativistic instability and forms a black hole, after which our code is unable to follow it. This model has the highest compactness of any solar-metallicity model in the \citet{swbj16} compilation and was widely believed not to explode \citep{2011ApJ...730...70O}. 

Interestingly, the 40-M$_{\odot}$ solar-metallicity model of \citet{swbj16} upon which we focus in this paper retained a hydrogen envelope of only $\sim$0.7 M$_{\odot}$. The binding energy of this envelope is correspondingly smaller. In fact, we find that, even including the binding energy of the entire mantle\footnote{$\sim$-2.5$\times$10$^{50}$ ergs exterior to 100,000 km; $\sim$-1.6$\times$10$^{51}$ ergs exterior to 2.5 M$_{\odot}$}, the explosion of this 40-M$_{\odot}$ progenitor has an energy near $\sim$1.6 Bethes.
The interval between the launch of the stalled shock and black hole formation in our model is $\sim$1.5 seconds, providing ample time for strong neutrino heating to power an explosion. Surprisingly, this suggests that the envelope will be ejected and a supernova explosion will accompany the formation of a relatively small black hole.   

The possibility that a solar-metallicity progenitor with such a high mass and compactness might explode vigorously, while still giving birth to a black hole, would represent a significant departure from current thinking. We present in this paper the hydrodynamic and physical results of our 3D F{\sc{ornax}} simulation of the 40-M$_{\odot}$ ZAMS star. First, in \S\ref{previous} we summarize a few of the salient precursor studies of relevance. In \S\ref{method}, we discuss the numerical specifications of the simulation and the basic physics incorporated into F{\sc{ornax}}. Then, in \S\ref{basics}, we present our overall hydrodynamic results
and place them into context. In \S\ref{signal}, we provide the neutrino ``light curves" and compare them with representative 3D results for lower-mass progenitors. We then transition in \S\ref{rotation} to a discussion of the the recoil kick and induced rotation of the core, and follow this in \S\ref{disk} with a description of the emergent disk.  In that section we also include a discussion and description of the neutrino-driven jets that issue from the rotating PNS. Next, in \S\ref{post} we summarize our 3D simulation using F{\sc{ornax}} of the post black hole formation phase.  In this simulation, we turn off neutrino transport, excise the core (leaving only a point gravitational mass), and follow the blast evolution for another $\sim$7 seconds to help determine the asymptotic state of the inner flow. Finally, we conclude in \S\ref{conclusions} with a synopsis of our findings and a general discussion of their import and limitations.

\section{Some Previous Theoretical Explorations}
\label{previous}

\citet{2020PhRvD.101l3013W} used their 3D code Prometheus-Vertex to simulate this same 40-M$_{\odot}$ model and did not witness an explosion.  However, they employed coarse angular ($\theta\, {\rm and}\, \phi$) resolution of 5$^{\circ}$ that has been shown to compromise the determination of explodability. In \citet{hiroki_2019}, we found that a model with an angular resolution of $\sim$2.8$^{\circ}$ failed to explode, while one with a resolution of $\sim$1.4$^{\circ}$ (our default) exploded and one with a resolution of $\sim$0.7$^{\circ}$ exploded even more vigorously. Suggestively, \citet{2020PhRvD.101l3013W} also simulated a $10^{-4}\times$solar metallicity 75-M$_{\odot}$ model, but at higher resolution ($\sim$0.9$^{\circ}$), with a correspondingly high compactness. That model's shock did explode, but due to the simultaneous prodigious mass infall onto the PNS, was quickly undermined into accretion by the rapid formation of a black hole within $\sim$100 milliseconds (ms).

Using the reduced-physics code 3D Coconut-FMT, \citet{chan2018} and \citet{moriya} did obtain an explosion of the 40-M$_{\odot}$ zero-metallicity progenitor \citep{2010IAUS..265....3W}, but the explosion, slightly artifically produced using an unphysical alteration in the charged-current absorption rate, was not deemed vigorous enough to avoid significant late-time fallback of much of the ejecta. This is due to the fact
that at zero-metallicity there is no wind mass loss and the progenitor retains a hydrogen-rich envelope, in this case of 24.7 M$_{\odot}$ with a helium core of 15.3 M$_{\odot}$. In general, zero- and low-metallicity progenitors are calculated to have both higher compactnesses (and, hence, infall accretion rates) and higher envelope binding energies for a given mass than the corresponding near-solar-metallicity massive stars. As a consequence, as \citet{2020PhRvD.101l3013W}, \cite{2018MNRAS.477L..80K}, and \citet{2023arXiv230706192K} have demonstrated, though such models might actually launch the stalled shock, black hole birth follows rapidly. The short time between explosion and black hole birth then either undermines the explosion or leads to a weak explosion due to the early truncation of neutrino heating.  This, coupled with the large mantle overburden, likely leads to an aborted explosion and the creation of a massive black hole through subsequent infall.

\adam{\citet{2021ApJ...914..140P} studied the same solar-metallicity 40-M$_{\odot}$ progenitor in 3D, but employed the quite approximate IDSA transport formalism, used a leakage scheme for the $\nu_{\mu}$ and $\nu_{\tau}$ neutrinos, ignored neutrino-electron and neutrino-nucleon energy redistribution, and employed the LS220 nuclear equation of state. They simulated rapidly-rotating, slowly-rotating, and non-rotating variants, found explosions for the two rotating models, but no explosion for the non-rotating model we study in this paper. They concluded that the slowly-rotating and non-rotating models left black holes (after 0.936 seconds and 0.776 seconds, respectively), but that the rapidly-rotating model did not leave a black hole. No model was carried beyond $\sim$1.0 seconds after bounce.} 

The work that most closely parallels ours here is that of \citet{ott2018_rel}. Those authors studied the same solar-metallicity 40-M$_{\odot}$ progenitor and used a sophisticated multi-group and general-relativistic 3D code {\it Zelmani} \citep{roberts:16} to explore it. They too observed the launch of the stalled shock. However, they simulated this 40-M$_{\odot}$ model to only $\sim$300 milliseconds after bounce and did not witness the formation of the black hole $-$ they merely surmised this outcome by extrapolation. Moreover, at the end of their simulation the blast had achieved an energy of only $\sim$0.1 Bethes, and this number did not include the negative mantle overburden energy. Furthermore, their code did not include velocity-dependent terms in the transport, nor the effects of neutrino energy redistribution due to electron and nucleon scattering.  The latter effect contributes to the neutrino heating in the gain region that drives the explosion.



\section{Method and Numerical Set-up}
\label{method}

For this study we use the F{\sc{ornax}} CCSN simulation code
\citep{skinner2019} which has been our workhorse for the investigation of core-collapse supernova explosions for the last seven years \citep{radice2017b,burrows2018,vartanyan2018a,vartanyan2018b,vsg2018,burrows_2019,radice2019,hiroki_2019,vartanyan2019,Nagakura2020,vartanyan2020,burrows_2020,nagakura2021,2021Natur.589...29B,2022MNRAS.510.4689V,coleman}.  Details concerning the numerical methodologies employed can be found in \citet{skinner2019} and in the appendix to \citet{vartanyan2019}. The transport algorithm solves for the zeroth and first (vector) moments of the transport equation in an M1 formulation with second- and third-moment analytic closures \citep{2023ApJ...943...78W}. The code is multi-group and velocity-dependent, and includes gravitational redshifts to the neutrino spectra.  The detailed neutrino-matter interaction suite is taken from \citet{burrows_thompson2004} and \citet{2006NuPhA.777..356B}, and nucleon-nucleon bremsstrahlung, weak magnetism and recoil corrections \citep{horowitz2002}, many-body corrections to the neutrino-nucleon scattering rates \citep{PhysRevC.95.025801}, and electron-capture rates on heavy nuclei \citep{2010NuPhA.848..454J} are all incorporated. Energy redistribution due both to neutrino-electron \citep{2006NuPhA.777..356B} and neutrino-nucleon scattering 
\citep{2020PhRvD.102b3017W} are included.  A multipole gravity solver to twelfth order is employed to calculate the 3D gravitational potential, and the dominant monopole term is given by the ``Case A" prescription of \citet{marek2006}. The latter embeds general-relativistic gravity in the form of a ``TOV"-like term into the hydrodynamics and hydrostatics and allows the code to sense the onset of the general-relativistic instability of the quasi-hydrostatic PNS core that leads to a black hole.

As stated in \S\ref{sec:int}, we use the solar-metallicity 40-M$_{\odot}$ progenitor model from \cite{swbj16}.
The outer boundary of the simulation is set at 100,000 kilometers (km) and the spherical grid was tiled with  1024$\times$128$\times$256 cells in radius, $\theta$, and $\phi$. We employed twelve neutrino energy groups spaced logarithmically between 1 and 300 MeV for the electron neutrinos ($\nu_e$) and between 1 and 100 MeV for the anti-electron ($\bar{\nu}_e$) and ``$\nu_\mu$" neutrinos (bundled as $\mu$ and $\tau$ neutrinos and anti-neutrinos).  We used the SFHo nuclear equation of state (EOS) \citep{2013ApJ...774...17S}, broadly consistent with extant theoretical and experimental constraints on the nuclear equation of state \citep{2023Parti...6...30L,2021PhRvL.126q2503R,2017ApJ...848..105T}. The progenitor model is initially non-rotating, though some degree of rotation is naturally induced due in part to continuing infall accretion \citep{coleman}, and as we will see (\S\ref{disk}) with this particular model-induced rotation is one of its many interesting features. The simulation was carried to the point of general-relativistic instability, after which the model and TOV gravity (sub)solver became unstable. The 3D models to which we for context compare the 3D 40-M$_{\odot}$ of this study were published in an earlier gravitational wave study (\citealt{vartanyan2023,burrows_2020}).

\section{Hydrodynamic Evolution of the 40-Solar-Mass Model}
\label{basics}

\begin{figure*}
    \centering
    \includegraphics[width=0.47\textwidth]{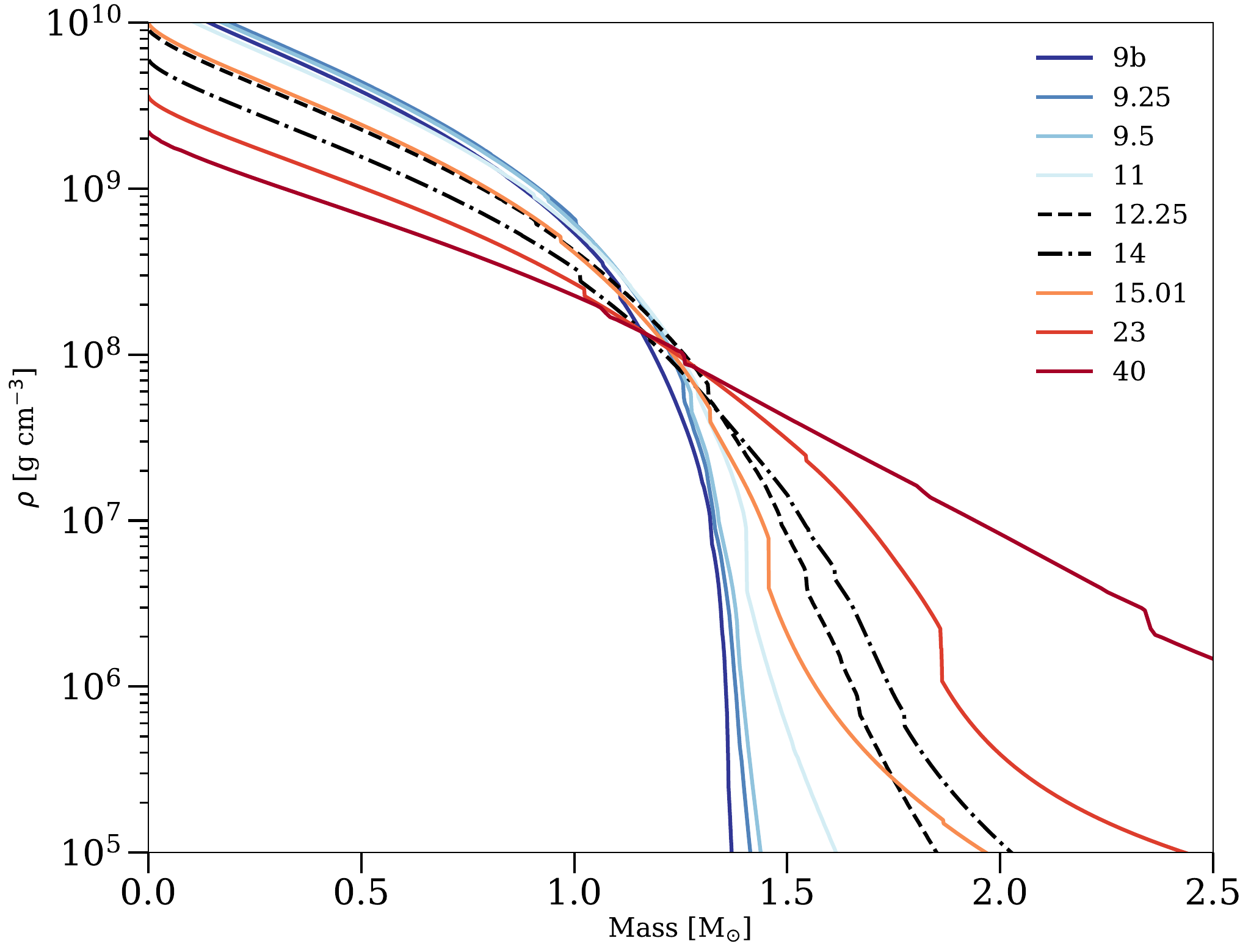}
    \includegraphics[width=0.47\textwidth]{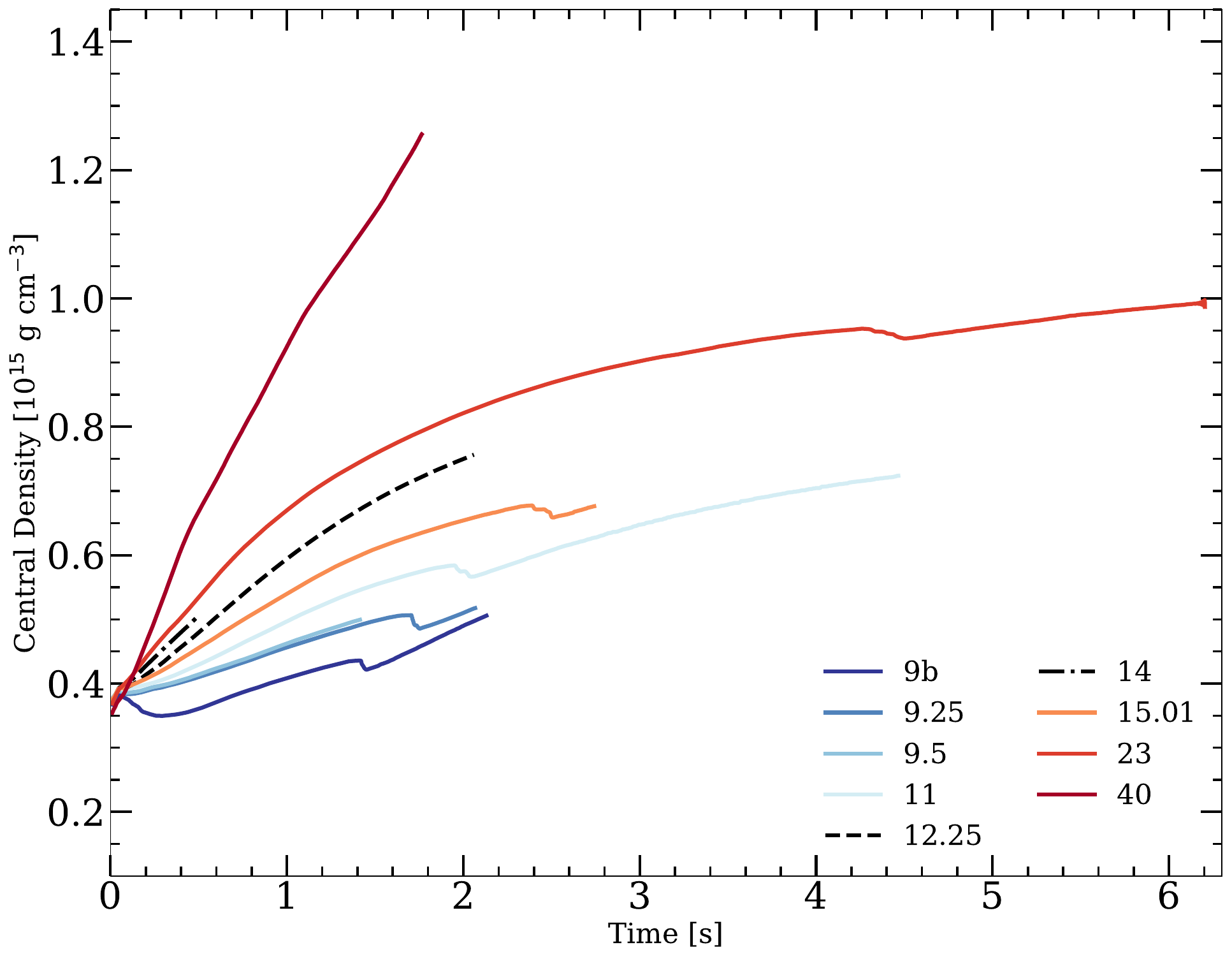}
    \caption{{\bf Left:} Density profiles (g cm$^{-3}$) versus enclosed mass (in M$_{\odot}$) for a collection of progenitor models. {\bf Right:} The evolution of the central baryon mass density for the models depicted in the left panel. Note the outlier behavior of the 40-M$_{\odot}$ model in both plots. The slight dip in central density seen for many of the models marks the time after bounce when PNS convection has reached the center.  Note that here the 9b model is distinguished from the 9a model, both published in \citet{vartanyan2023}, by the lack of imposed initial perturbations in the former.}
    \label{fig:rho_full}
\end{figure*}

The left-hand panel of Figure\,\ref{fig:rho_full} depicts the initial mass density profiles in the stellar core for a representative collection of massive-star solar-metallicity progenitor models from \citet{swbj16} and \citet{sukhbold2018} at the time of collapse. Note the outlier nature of the 40-M$_{\odot}$ ZAMS mass progenitor.  Its shallower density profile translates into significantly higher infall accretion rates, higher accretion-powered neutrino luminosities after bounce, and the rapid accumulation of mass in the PNS core. The silicon/oxygen interface for this model is seen at $\sim$2.35 M$_{\odot}$, significantly further out than for most models.  Also, shown in black are two models (the 12.25- and 14-M$_{\odot}$ models) that \citet{wang} and \citet{vartanyan2023} have previously found do not explode and that will form black holes by a different channel (see \S\ref{conclusions}). These models are at intermediate compactness and have weak silicon/oxygen density and entropy jumps, structures that have difficulty exploding by the neutrino mechanism \citep{wang}. 

The corresponding evolution after bounce of the central mass densities for the associated 3D full-physics collapse simulations are provided in the right-hand panel of Figure\,\ref{fig:rho_full}. The rapid relative evolution in the central density of the 40-M$_{\odot}$ model is apparent, which by the time of the general-relativistic instability at $\sim$1.76 seconds after bounce has reached almost 1.3$\times$10$^{15}$ g cm$^{-3}$. The slight decreases of the central density in the other models  at from $\sim$1.6 to $\sim$4.3 seconds for from smaller to larger ZAMS-mass progenitors flag the times that PNS convection reaches the center of the core \citep{nagakura_pns}. This never happens for the 40-M$_{\odot}$ model.  

\begin{figure}
    \centering
    \includegraphics[width=0.47\textwidth]{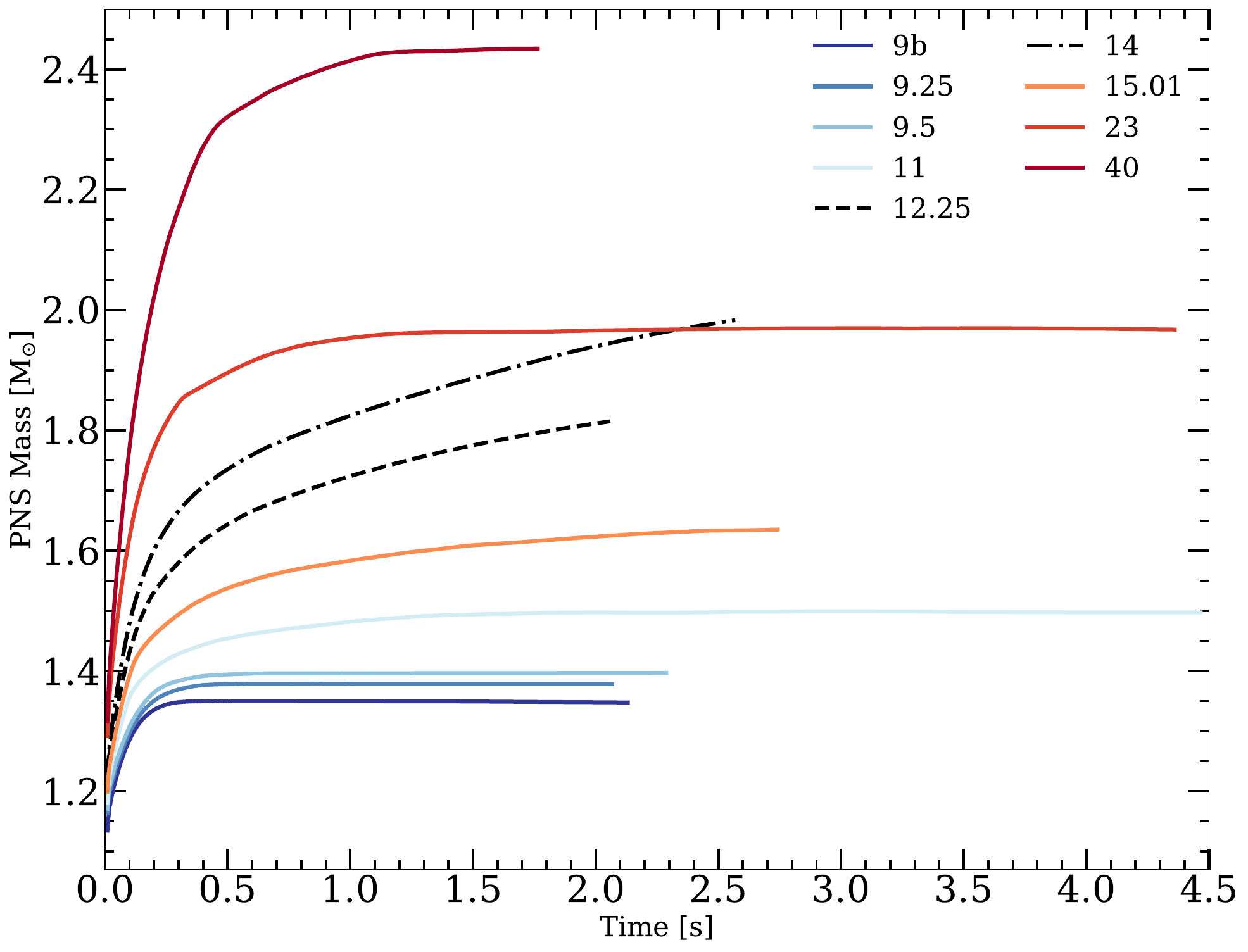}
    \caption{Baryonic PNS mass in M$_{\odot}$ as function of time after bounce (in seconds) for a collection of 3D CCSN models. Note the outlier behavior of the 40-M$_{\odot}$ model, due to its relatively high progenitor compactness and correspondingly high mass infall/accretion rate.}
    \label{fig:pns}
\end{figure}

Figure\,\ref{fig:pns} portrays the growth of the PNS baryon mass versus time after bounce for the 40-M$_{\odot}$ model and the comparison model suite. Again, the growth of the PNS mass of the 40-M$_{\odot}$ model is quite rapid. At the time of black hole formation, the baryon mass of the PNS core is 2.434 M$_{\odot}$ and its gravitational mass is 2.298 M$_{\odot}$.  However, at the launch of the shock, the PNS baryon mass is $\sim$2.1 M$_{\odot}$. \adam{We emphasize that the maximum gravitational mass in fact depends upon the entropy and composition (Y$_e$) profile and is not a single number, depending as it does upon evolution.  The standard quoted gravitational mass is for the "cold/catalyzed" state in beta equilibrium between n, p, and e$^-$ for the neutrino-transparent state.  The maximum gravitational and baryonic masses for this asymptotic state and the SFHo EOS are 2.06 M$_{\odot}$ and 2.42 M$_{\odot}$, respectively \citep{2013ApJ...774...17S}.} 

Though the accretion rate onto the PNS is near $\sim$1.0 M$_{\odot}$ s$^{-1}$ at the onset of explosion, it is near $\sim$10$^{-2}$ M$_{\odot}$ s$^{-1}$ when the black hole eventually forms.  Much of the matter that encounters the shock is in fact ejected, but a fraction is merely slowed down and deflected, and continues inward. That subset results in the accretion onto the core of infalling plumes predominantly in one or a few angular sectors, while the explosion proceeds in others. Frequently, the neutrino-driven explosion starts with a dipolar instability that accompanies the monopolar instability that is the supernova explosion \citep{dolence_2015,burrows2018,burrows_2020,wang}. This behavior is generic in 3D CCSN simulations $-$ there is simultaneously accretion and explosion, something not possible in 1D. These infalling plumes, generally with lower entropies and higher densities than found in the surrounding ejecta, accrete onto the core over a spread of times. They are also responsible for the induced spin-up of the PNS residue (\S\ref{rotation}) and the excitation of oscillatory modes of the PNS that radiate gravitational radiation \citep{radice2019,vartanyan2023}. Interestingly, a fraction of the matter around the silicon/oxygen interface, which starts infalling spherically, accretes onto the core aspherically over a spread in time of a few hundred milliseconds. This spread is reflected in the slope changes in PNS mass from $\sim$0.4 to $\sim$0.6 seconds seen for the 40-M$_{\odot}$ model in Figure \ref{fig:pns}. Since the explosion starts near $\sim$0.25 seconds, the accretion of the silicon/oxygen interface, so often a factor in kick-starting CCSN explosions, is not a factor in inaugurating this explosion.

Importantly, the explosion we witness in our simulation of the 40-M$_{\odot}$ star occurs near $\sim$0.25 seconds after bounce, after which as much $\sim$0.35 M$_{\odot}$ is still accreted before the black hole forms.  This is a delay between explosion and black hole formation of $\sim$1.5 seconds, during which the integrated neutrino energy deposition in the ``gain region" \citep{1985ApJ...295...14B} accumulates. If the delay between explosion and black hole formation were of short duration, the total energy deposition would be quite small, making the expulsion of most of the star impossible and ensuring the eventual formation of a massive black hole without a canonical generic supernova light curve.  This is what \citet{2020PhRvD.101l3013W}, \cite{2018MNRAS.477L..80K}, \citet{2023arXiv230706192K}, and \citet{ott2018_rel} found. As an aside, we note that we have simulated the same model in 1D.  It forms a black hole near $\sim$0.567 seconds when its baryon mass is $\sim$2.407 M$_{\odot}$.

\begin{figure}
    \centering
    \includegraphics[width=0.47\textwidth]{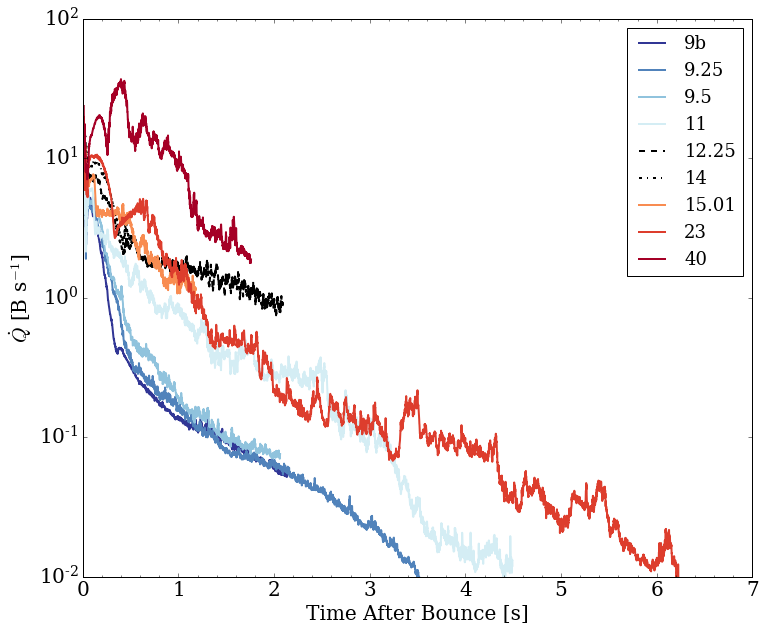}
    \caption{Net neutrino heating rate in the gain region as a function of time for a collection of comparison 3D models. Note that the 40-M$_{\odot}$ model experiences a significantly higher heating rate than any other model. \adam{Note also that this plot
    does not show the degree of cooling interior to the gain region, but exterior to the inner core and neutrinospheres. This quantity is generally large for those models such as the 14- and 12.25-M$_{\odot}$ models shown here for which the stalled shock recedes significantly and which follows one suggested channel of black hole formation.}}
    \label{fig:Qdot}
\end{figure}

Figure \ref{fig:Qdot} depicts this heating history for the 40-M$_{\odot}$ model and various comparison 3D models. This figure demonstrates the large net heating rate and power that drives the explosion of our 40-M$_{\odot}$ model. This high rate is a consequence of the singularly high compactness and accretion rates of this star. The long duration to ``second collapse" and high net heating rate
ensure a vigorous outcome. The formation of the black hole abruptly truncates this driving neutrino power, but at that time the total explosion energy (internal, gravitational, and kinetic), including all the overburden of the entire star, is $\sim$1.416 Bethes (but see \S\ref{post}).

\begin{figure}
    \centering
    \includegraphics[width=0.47\textwidth]{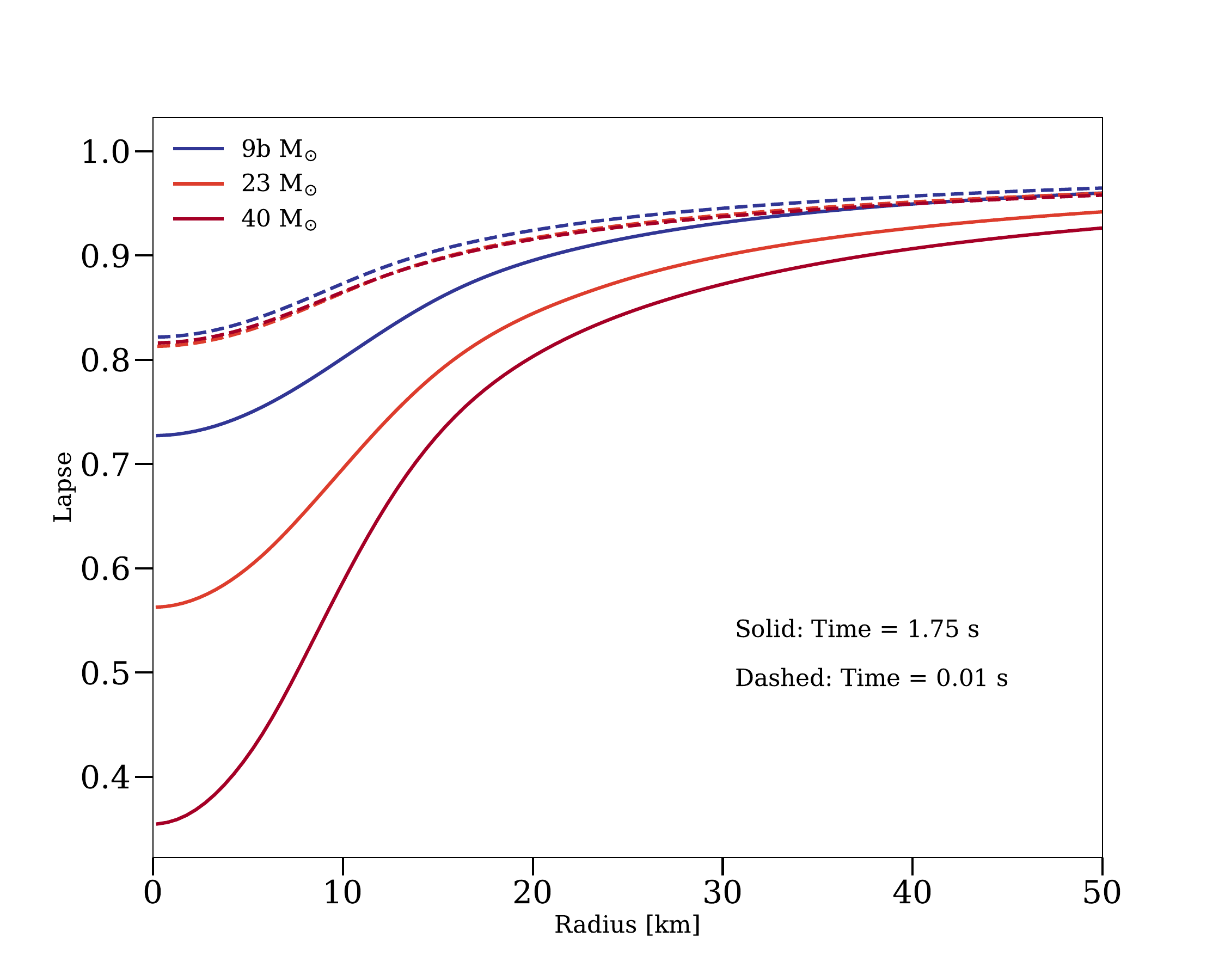}
    \caption{Here, we plot the mean lapse radial profiles at 0.01 seconds after bounce and 1.75 seconds after bounce for our 3D 9b, 23, and 40 M$_{\odot}$ models. Notice how much more deeply the 40-M$_{\odot}$ profile plunges, achieving a value in the center of $\sim$0.35 at the time of black-hole formation ($\sim$1.76 seconds after bounce). See text for a discussion.}
    \label{fig:lapse}
\end{figure}

In Figure\,\ref{fig:lapse}, we plot the radial profiles of the lapse function at 0.01 seconds after bounce and 1.75 seconds after bounce for our 3D 9b, 23, and 40 M$_{\odot}$ progenitor models. The lapse is the standard measure of the gravitational redshift and the strength of general-relativistic corrections to the hydrodynamics. The 40-M$_{\odot}$ profile plunges much more deeply, achieving a value in the center of $\sim$0.35 at the time of black-hole formation. This is very similar in value to the central lapse achieved in our 1D simulation of the same star and to the lapses generally found at instability in \citet{2011ApJ...730...70O}. Variations are expected in the precise central lapse, baryon mass, and gravitational mass at black hole formation due to variations in the entropy, electron fraction, and angular velocity profiles, as well as (and importantly) the nuclear equation of state (see \S\ref{conclusions} for a discussion). 

\begin{figure*}
    \centering
    \includegraphics[width=0.6\textwidth]{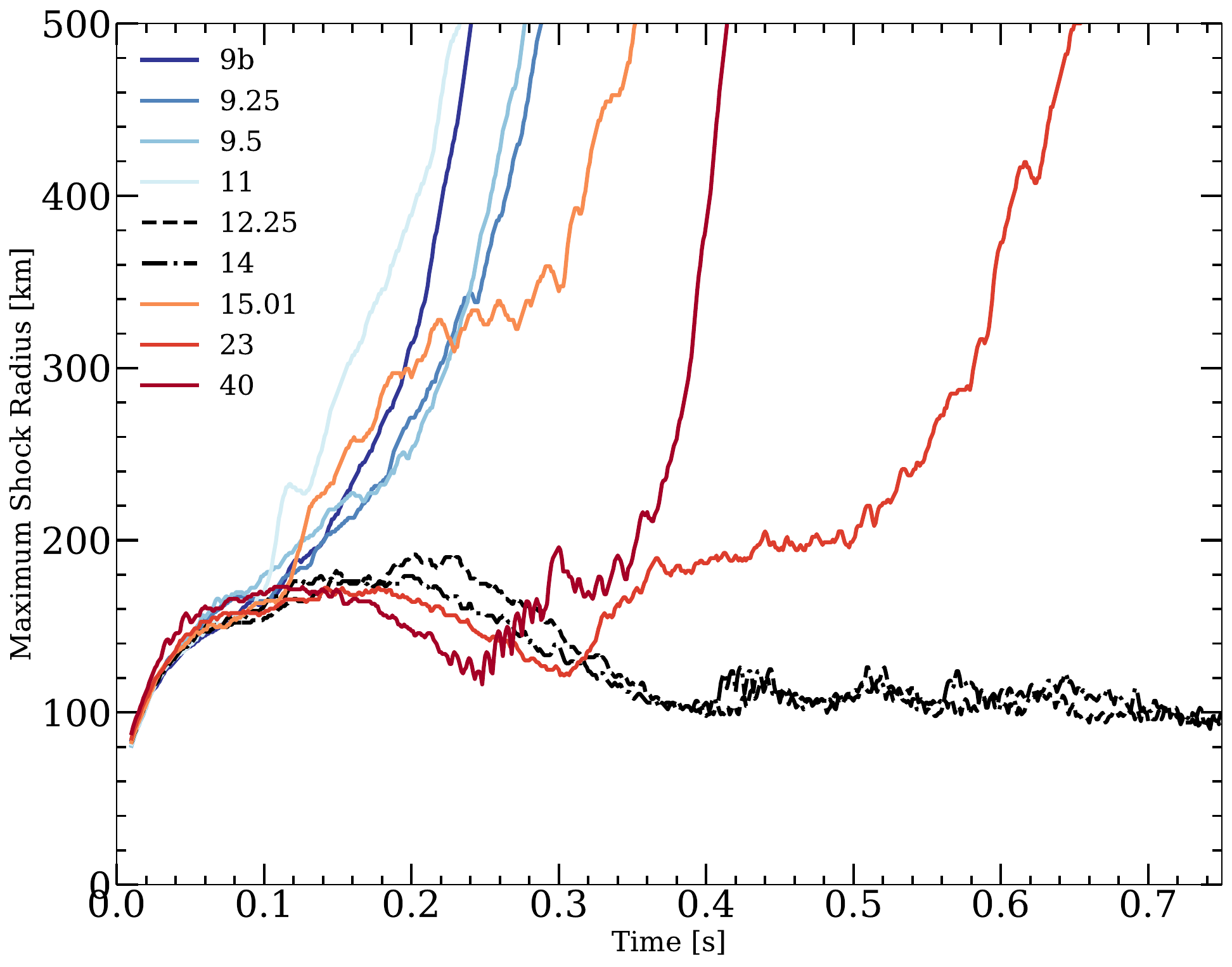}
    \includegraphics[width=0.6\textwidth]{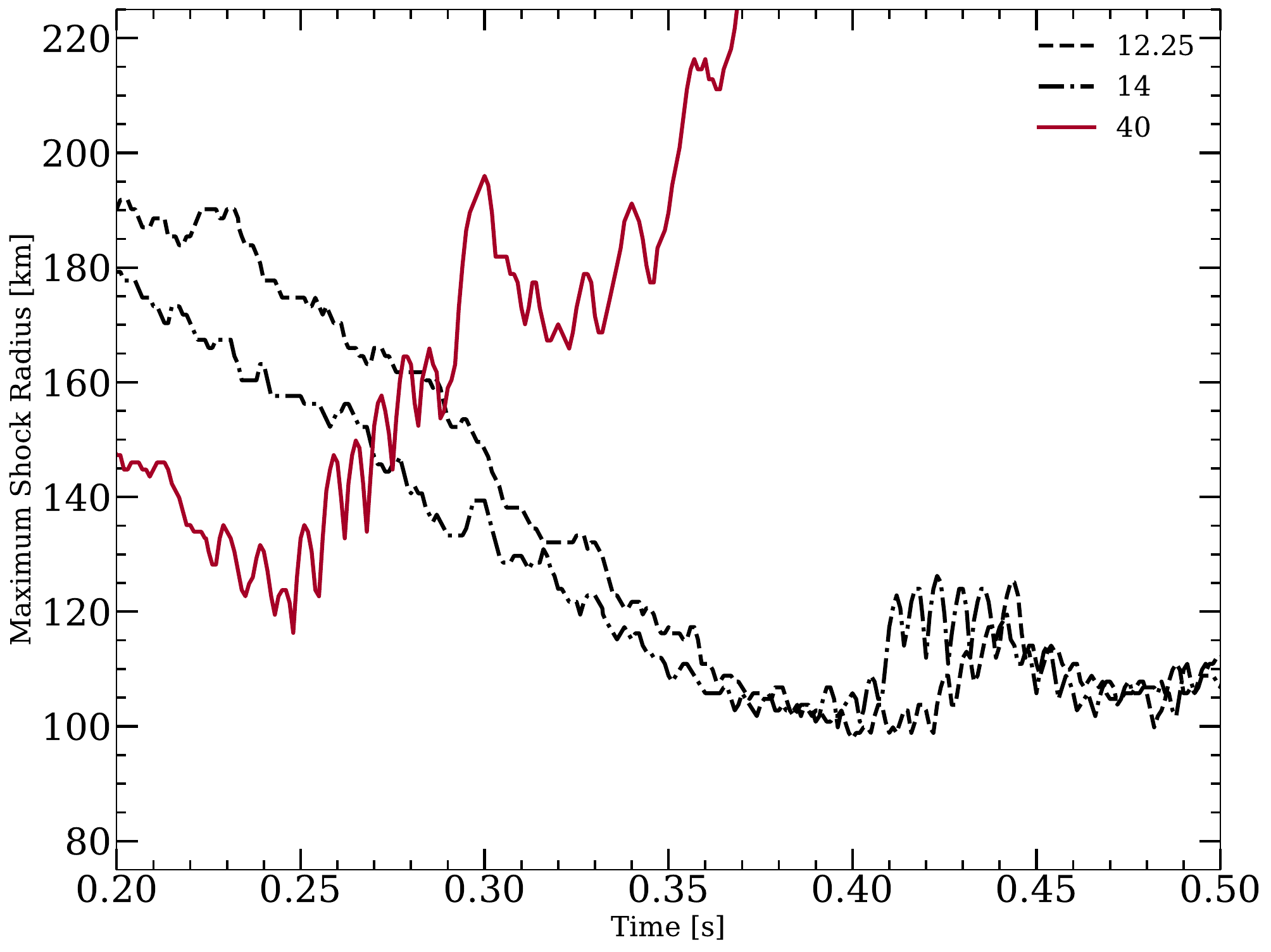}
    \caption{{\bf Top:} Maximum shock radii (in kilometers, km) as a function of time after bounce (in seconds) for a collection of 3D models. The black lines trace the evolution of the two models shown here (the 12.25- and 14-M$_{\odot}$ ZAMS progenitors) that do not explode by the neutrino mechanism and are destined to form black holes. Note that the maximum shock radius for the 40-M$_{\odot}$ model first decreases, then starts to experience a combination of the SASI and Spiral-SASI over a period of $\sim$100 ms before and as it explodes. {\bf Bottom:} A closed-up of the left-hand plot provided to focus in on the spiral SASI-like oscillations as seen via the motion of the shock radius for the three models shown here that form black holes, though by two different mechanisms. The SASI class oscillations for the two non-exploding black-hole formers can be seen starting near $\sim$0.4 seconds and continue thereafter. See the text for a discussion. }
    \label{fig:rs}
\end{figure*}

\begin{figure}
    \centering
    \includegraphics[width=0.47\textwidth]{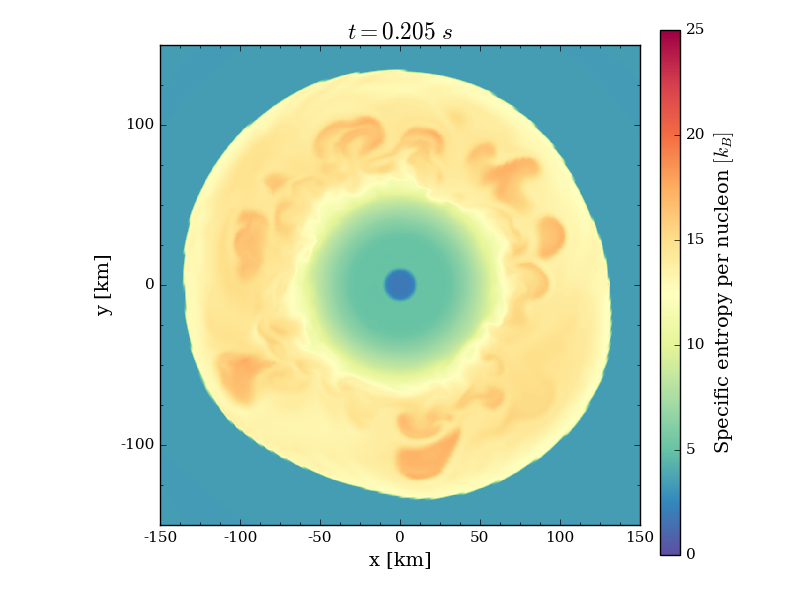}
    \includegraphics[width=0.47\textwidth]{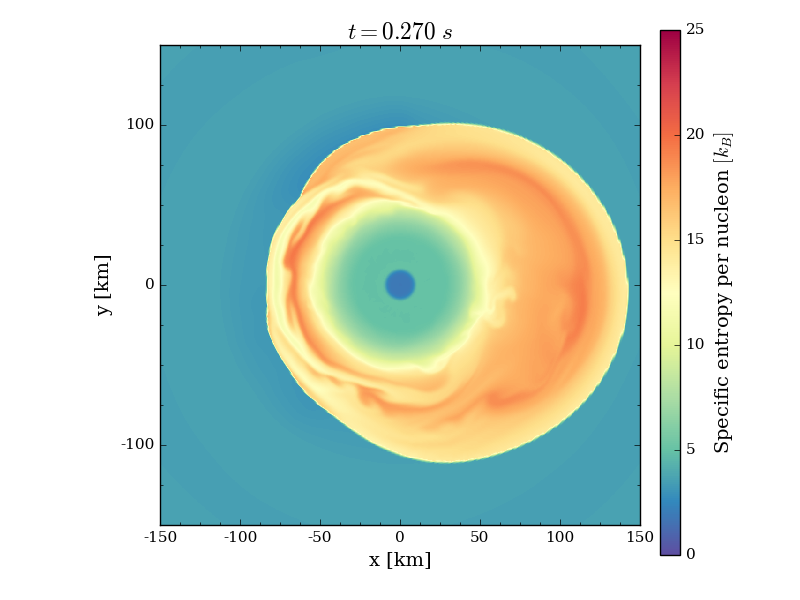}
    \includegraphics[width=0.47\textwidth]{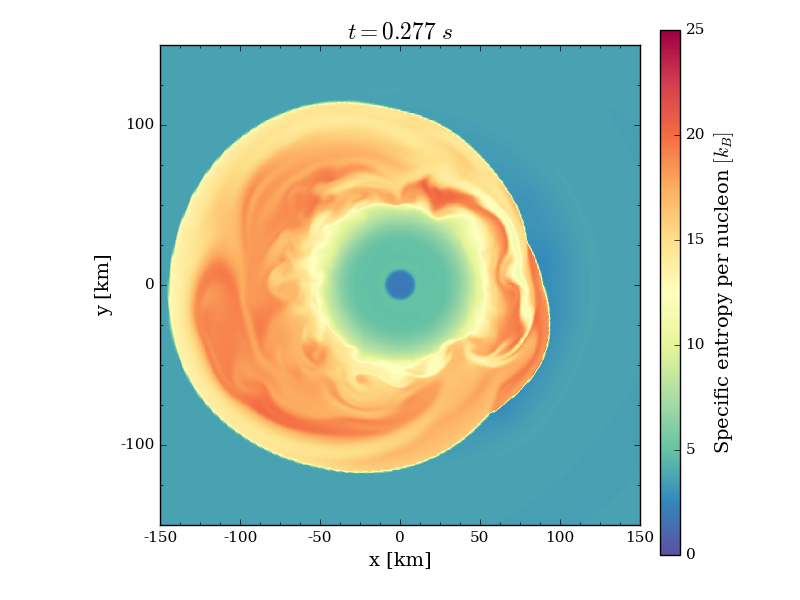}
    \caption{Entropy map before (\textbf{top}) and during (\textbf{middle and bottom}) the excitation of  the SASI/Spiral-SASI mode.}
    \label{fig:sasi}
\end{figure}

The top panel of Figure\,\ref{fig:rs} depicts the maximum shock radii during the early post-bounce phase for 3D simulations of a sample of solar-metallicity models with a representative spread of ZAMS masses, most of which explode. This includes the 40-M$_{\odot}$ model in red. However, two of the models (12.25- and 14-M$_{\odot}$, in black) do not and are expected to form black holes after tens of seconds or tens of minutes (see \citealt{vartanyan2023}). However, as the maximum shock radii of all three models initially decrease and achieve values for which explosion is not generally expected, a predominantly spiral SASI (``Standing-Accretion Shock Instability") \citep{blondin:03,blondin_shaw} mode erupts in all three. This is expected for such small accretion shock radii and we suggest is a common feature of black hole formation \citep{2020PhRvD.101l3013W,vartanyan2023}. For the 12.25- and 14-M$_{\odot}$ models, the shocks settle into steady spiral SASI oscillations at a period given roughly by the matter advection time to the core ($\sim$10 milliseconds) \citep{foglizzo2002,foglizzo:07}. These are also modulated by a $\sim$70-millisecond ``breathing mode," seen in this top panel and in \citet{vartanyan2023}. However, soon into its spiral SASI oscillation, the stalled shock of the 40-M$_{\odot}$, energized by its high net neutrino heating power, starts to increase, even as it continues to execute the spiral SASI. This is more clearly seen in the bottom panel of Figure\,\ref{fig:rs}. We suggest that the periodic breathing modulation seen in the 12.25- and 14-M$_{\odot}$ models is a stable oscillation driven by neutrino heating and cooling that in the 40-M$_{\odot}$ model is actually unstable and growing, eventually transitioning into an explosion. This seems to be what we are seeing. The early SASI motion of the 40-M$_{\odot}$ model is captured in Figure \ref{fig:sasi}.

In Figure\,\ref{fig:ent}, on a large scale
20,000 km on a side, we show three perpendicular planar slices of the entropy just before black hole formation.
The position of the inner PNS core from which the black hole will emerge within $\sim$10 milliseconds of these snapshots is artificially depicted with a black dot. The inner blue/light-blue interface is the neon/oxygen interface in the progenitor. Clearly shown, in particular with the y-x slice, is the significant degree of anisotropy of the explosion, as well as of the entropy distribution.  In fact, neutrino-driven jets (\S\ref{disk}), emerging in roughly opposite directions, are seen to dominate the late-time flow. The speed of some of the matter in these jets reaches $\sim$45,000 km s$^{-1}$. At this time, the shock wave (outer light-blue/white interface) has reached $\sim$20,000 km in one direction, while in another it has achieved only $\sim$10,000 km.  In the top right plot, the ``shadow" region near the core at ``5 o'clock" contains matter still infalling. The total baryon mass of this infalling matter is between 0.1 and 0.2 M$_{\odot}$ $-$ we show in \S\ref{post} that after seven more seconds of evolution the additional mass joining the black hole is in fact $\sim$0.2 M$_{\odot}$ and that the final baryon mass of the black hole will be $\sim$2.63 M$_{\odot}$ at $\sim$8.8 seconds after bounce. We emphasize that this infalling matter is not ``fallback," since none of this matter ever achieved positive radial velocities. The shock wave swept through it, but did not reverse its velocity and at the time of these stills it is falling inward. We return to a discussion of the 3D structures that emerge in the interior, in particular the neutrino-driven jets and disk, in \S\ref{disk}.




\section{Neutrino and Gravitational-Wave Emissions in Context}
\label{signal}

The signatures, in particular the neutrino and gravitational-wave emissions, are important potential diagnostics of the event. Moreover, the neutrinos, as agents of explosion, play a central role overall in the CCSN mechanism. In Figure\,\ref{fig:lum}, we compare the angle-averaged neutrino luminosities of the 40-M$_{\odot}$ model with those for a small sample of representative 3D models. The top, middle, and bottom panels are for the $\nu_e$, $\bar{\nu}_e$, and ``$\nu_{\mu}$" neutrinos, respectively. As stated, two of these models did not explode and will likely form black holes much later. This plot is to be compared with the corresponding plot for the net neutrino heating power in Figure \ref{fig:Qdot}. 

Figure \ref{fig:lum} clearly shows the relatively high neutrino luminosities for the 40-M${_\odot}$ progenitor. Between $\sim$0.4 and $\sim$0.6 seconds, the accretion over time of the silicon/oxygen interface results in the drop in luminosity seen, after which steady, though diminishing, accretion powers the neutrino emissions. Note that after $\sim$1.2 seconds, the neutrino luminosities of the two other black hole formers exceed that of the 40-M$_{\odot}$. This is because at this time the accretion rate onto the PNS cores of the non-exploding 12.25- and 14-M$_{\odot}$ models exceeds that of the exploded 40-M$_{\odot}$ model.  

In Figure\,\ref{fig:gw}, we show the two polarizations of the gravitational wave strains along the x-direction for the 9b-, 23-, and 40-M$_{\odot}$ models. The huge contrast between the 40-M$_{\odot}$ and 9b-M$_{\odot}$ signals is startlingly plain to see. This is made all the more clear in Figure\,\ref{fig:Egw_matter}, which indicates with the left panel the factor of $\sim$4$\times$10$^{3}$ spread in total radiated gravitational-wave energy between these two models. In the right panel, a similarly wide range in the contrast of the gravitational-wave emission off the mass-energy of the emergent neutrinos is seen for the entire suite of representative models, with the 40-M$_{\odot}$ clearly the most ``luminous."  The relatively large neutrino component of the radiated gravitational-wave energy of the 40-M$_{\odot}$ model reflects the large asphericity of the associated neutrino emissions. Also, of interest is the large matter memory (strain offset) seen in Figure \ref{fig:gw} for the 40-M$_{\odot}$ model that reflects the significant ejection asymmetry (see Figure \ref{fig:ent}) in the blast itself.


\section{Recoil Kick and Induced Rotation}
\label{rotation}

The anisotropic blast hydrodynamics, aspherical infall, asymmetric neutrino emissions, and integrated anisotropic gravitational attraction between the core and clumpy ejecta together result in a net recoil kick of the PNS core. Pulsars as a group manifest some of the fastest speeds seen in the galaxy \citep{Lyne1994,Ng2007,faucher_kaspi,Yang2021,burrows1996,Scheck2006,scheck2008,Nordhaus2012,Wongwathanarat2013,Janka2017,coleman}, but black holes have generally been thought to receive a significantly smaller recoil kick \citep{2005ApJ...625..324W,2019MNRAS.489.3116A,oh.bh.kick,coleman}. However, the grossly asymmetrical explosion dynamics
of our 40-M$_{\odot}$ model results in a kick speed (absolute value of the vector sum of the various components) just before black hole formation of $\sim$1150 km s$^{-1}$ \citep{Janka2013}, using the formalism of \citet{coleman}.  \adam{The arrows on Figure \ref{fig:ent} indicate the kick direction at the time of black hole formation. This direction is opposite to the direction in which most of the matter is ejected.}
%
%
The absolute value of the neutrino kick is subdominant, but reaches $\sim$400 km s$^{-1}$. The aggregate kick speed is quite unlike the birth kicks inferred for the stellar-mass black holes in X-ray binaries \citep{2005ApJ...625..324W,2019MNRAS.489.3116A,oh.bh.kick} and suggests that the black holes of this birth channel are unlikely to have companions. This would make observing the population of such black holes exceedingly difficult. Nevertheless, this is what we find.

The induced spin of the initially non-rotating core results from the accretion of the infalling plumes of lower-entropy, higher-density matter (see \S\ref{basics}) in a roughly stochastic fashion. This is a generic feature of core-collapse physics in 3D \citep{coleman}.  \adam{During the proto-neutron star phase, its surface is arbitrarily and traditionally defined as the 10$^{11}$ g cm$^{-3}$ isosurface.} The last phases of such accretion in this 40-M$_{\odot}$ progenitor context lead to a ring of matter with only a few hundreds of a solar mass rotating in roughly a disk structure (see \S\ref{disk}). The angular momentum resides only in a dense rotating disk of matter at the periphery of the PNS core. The deep core is not spun up in our simulations. 
%
%
At the end of our full-transport simulation at 1.76 seconds after bounce, just at black hole formation, the specific angular momentum of the core is $\sim$1.77$\times$10$^{14}$ cm$^2$ s$^{-1}$. This number is derived by artificially distributing the measured $\sim$8.5$\times$10$^{47}$ (cgs) of angular momentum into the entire core and is done to compare with the enabling specific angular momentum ($\sim$2.0$\times$10$^{16}$ cm$^2$ s$^{-1}$ \adam{$\equiv$ $\sim$1$\times$10$^{50}$ cgs total}) of long-soft gamma-ray bursts (GRBs) in the collapsar model \citep{1993ApJ...405..273W}. At this stage, we do not find a total angular momentum sufficient for the collapsar model. However, subsequent accretion (see \S\ref{post}) over the next $\sim$7 seconds boosts this up by a factor of $\sim$seven. The resulting relativistic spin parameter ``a" (in units of $GM^2/c$) is $\sim$0.19, one-fifth of the maximum possible. Clearly, the possibility of an MHD-driven jet post-collapse is intriguing. Unfortunately, our code does not yet boast magnetic fields. \adam{We note that the ``final" spin for this channel of black hole formation is generally larger than those inferred for the compact objects in observed black-hole binaries ($\sim$0.01-0.05)}\citep{2006ARA&A..44...49R,2009ApJ...697..900M}. 




\section{Emergence of a Disk and Neutrino-Driven Jets}
\label{disk}

\begin{figure*}
    \centering
    \includegraphics[width=0.49\textwidth]{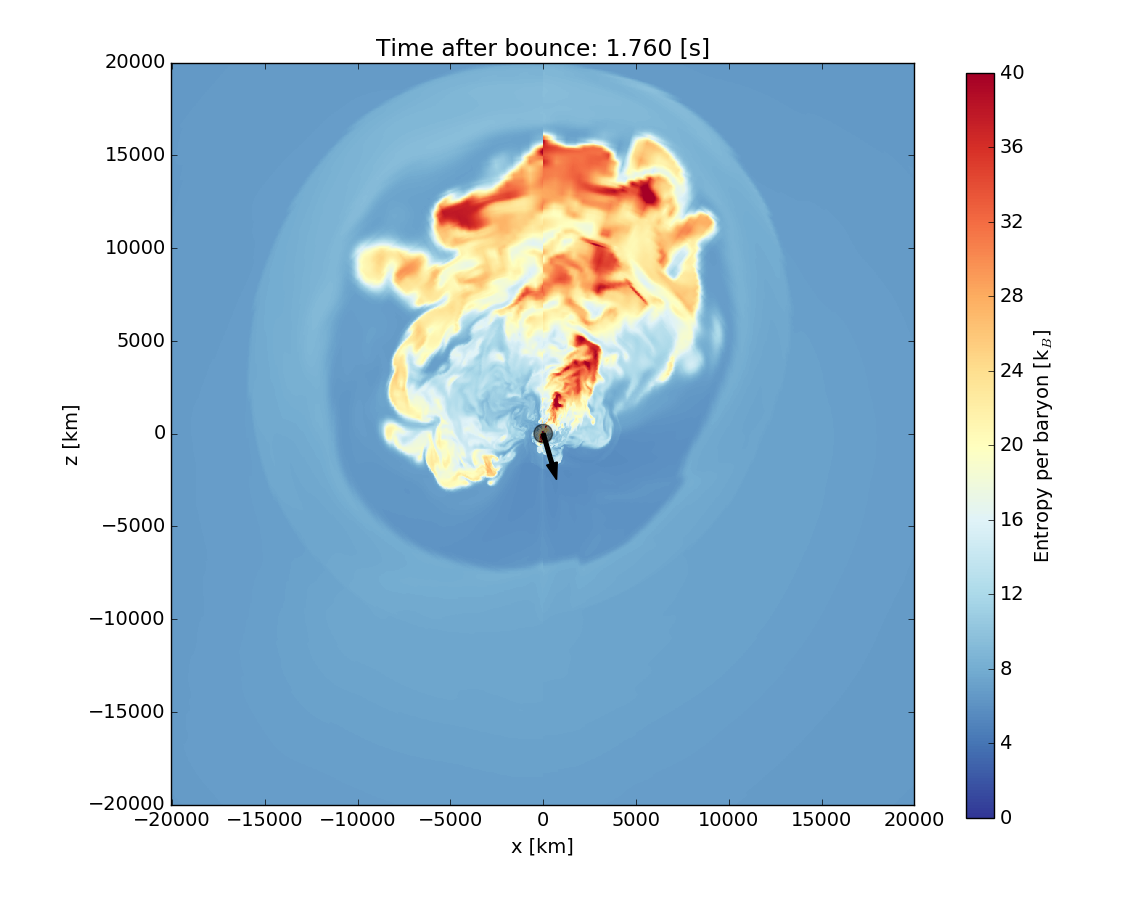}
    \includegraphics[width=0.49\textwidth]{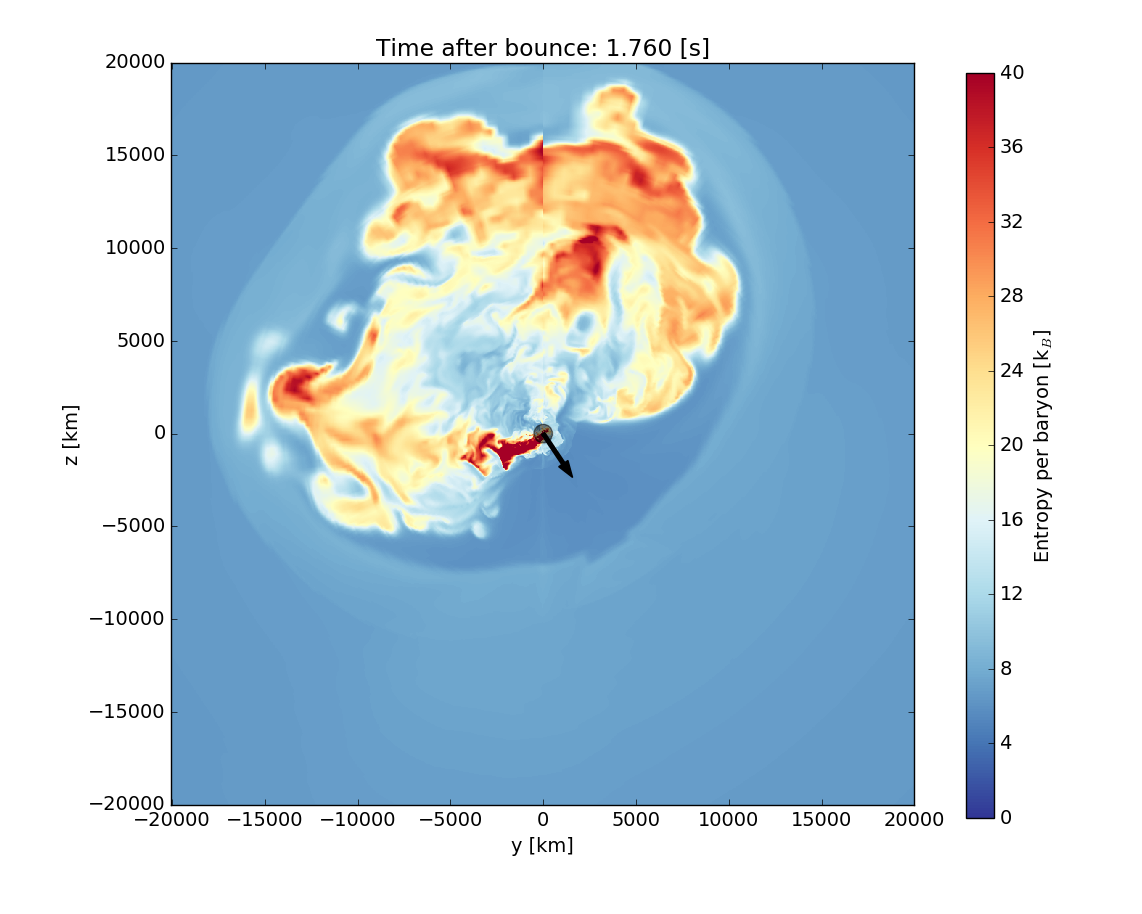}
    \includegraphics[width=1.0\textwidth]{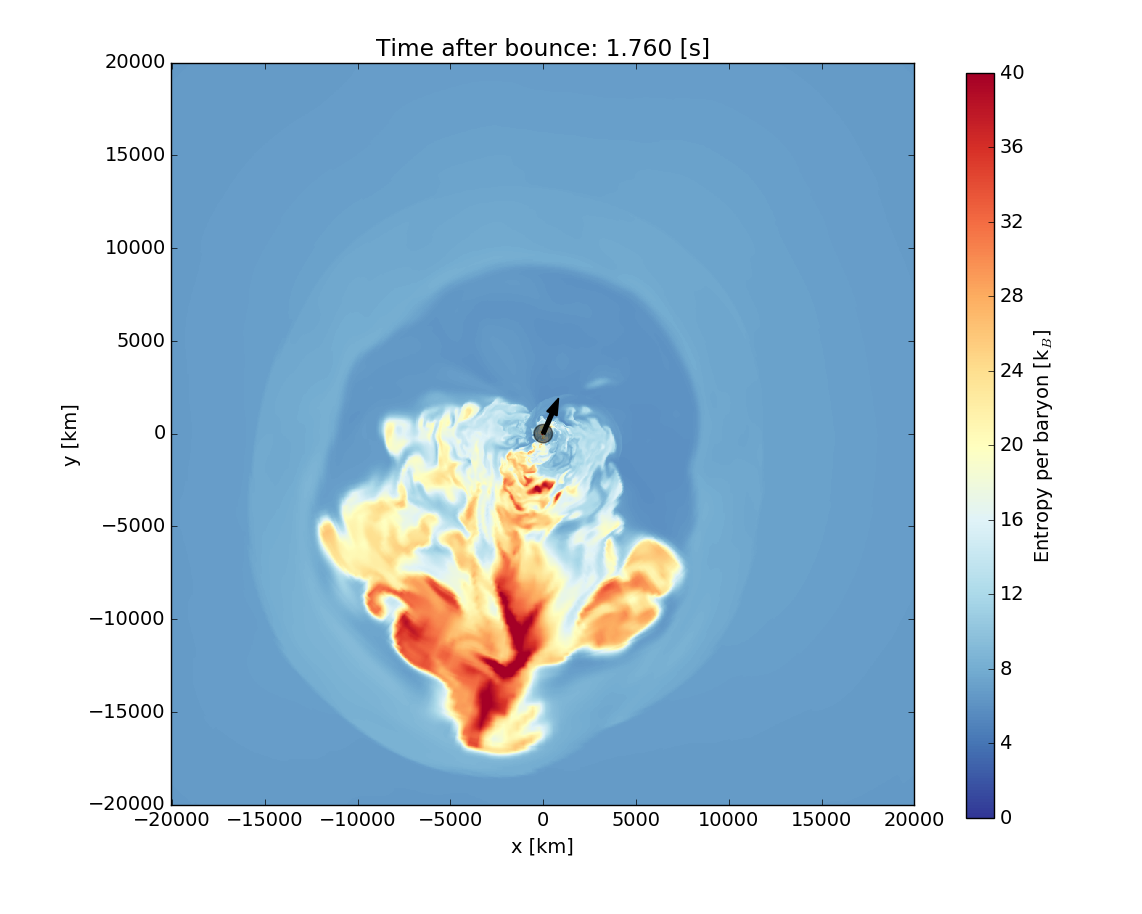}
    \caption{Entropy slices at 1.76 seconds post-bounce for the 40-M$_{\odot}$ solar-metallicity progenitor in the z-x, z-y, and y-x planes. The black circle shows the position of the central object, while the arrows show the projected kick velocity directions. Note the very asymmetric morphology (most notable in the y-x plane) and ongoing significant infall accretion (most notable in the z-y plane,  at ``five o'clock").  The two high-entropy jets originating from the central compact object and their complicated cocoon structures are clearly in evidence (in red). The inner blue/light-blue interface is the neon/oxygen interface in the progenitor. See text in \S\ref{basics} and \S\ref{disk} discussions.}
    \label{fig:ent}
\end{figure*}

\begin{figure*}
    \centering
    \includegraphics[width=0.7\textwidth]{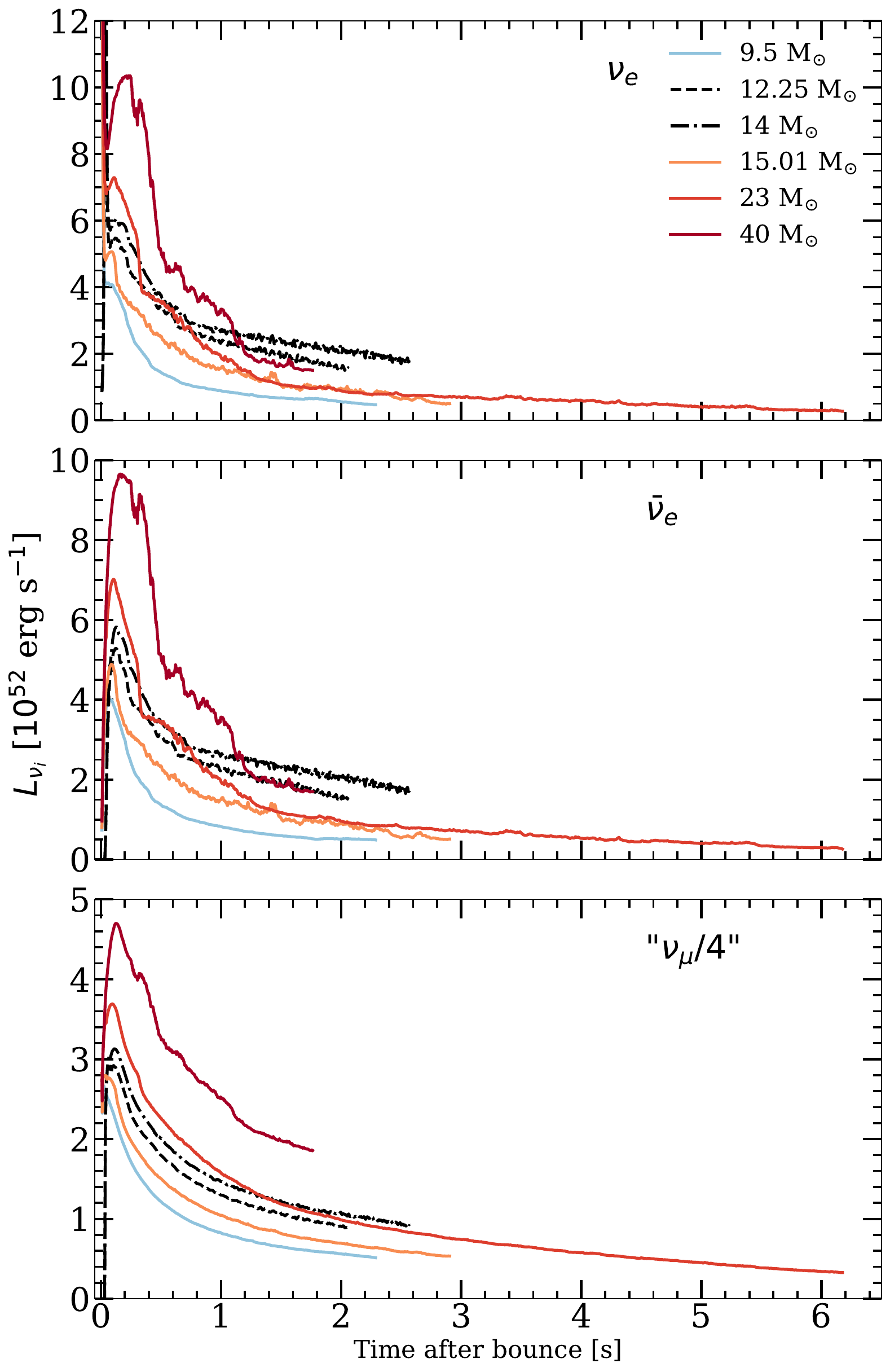}
    \caption{Angle-averaged neutrino luminosities in the observer frame for the $\nu_e$, $\bar{\nu}_e$, and ``$\nu_{\mu}$" species at 500 km as a function of time after bounce. The 40-M$_{\odot}$ model shows the aforementioned outlier behavior, with its significantly higher neutrino luminosity for all species, corresponding to its higher compactness and accretion rate. The black lines are for the 12.25- and 14-M$_{\odot}$ black hole formers.}
    \label{fig:lum}
\end{figure*}

\begin{figure}
    \centering
    \includegraphics[width=0.47\textwidth]{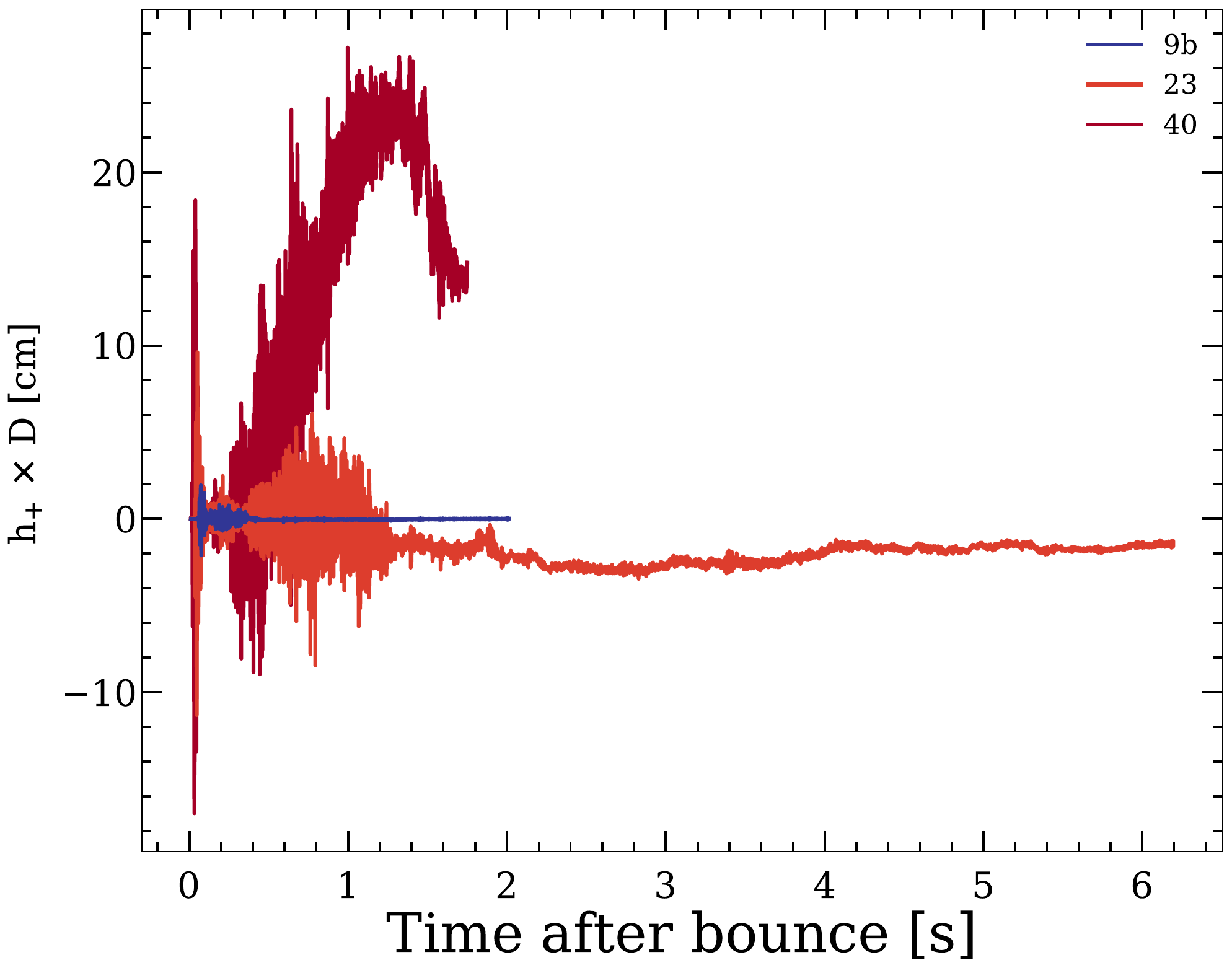}
    \includegraphics[width=0.47\textwidth]{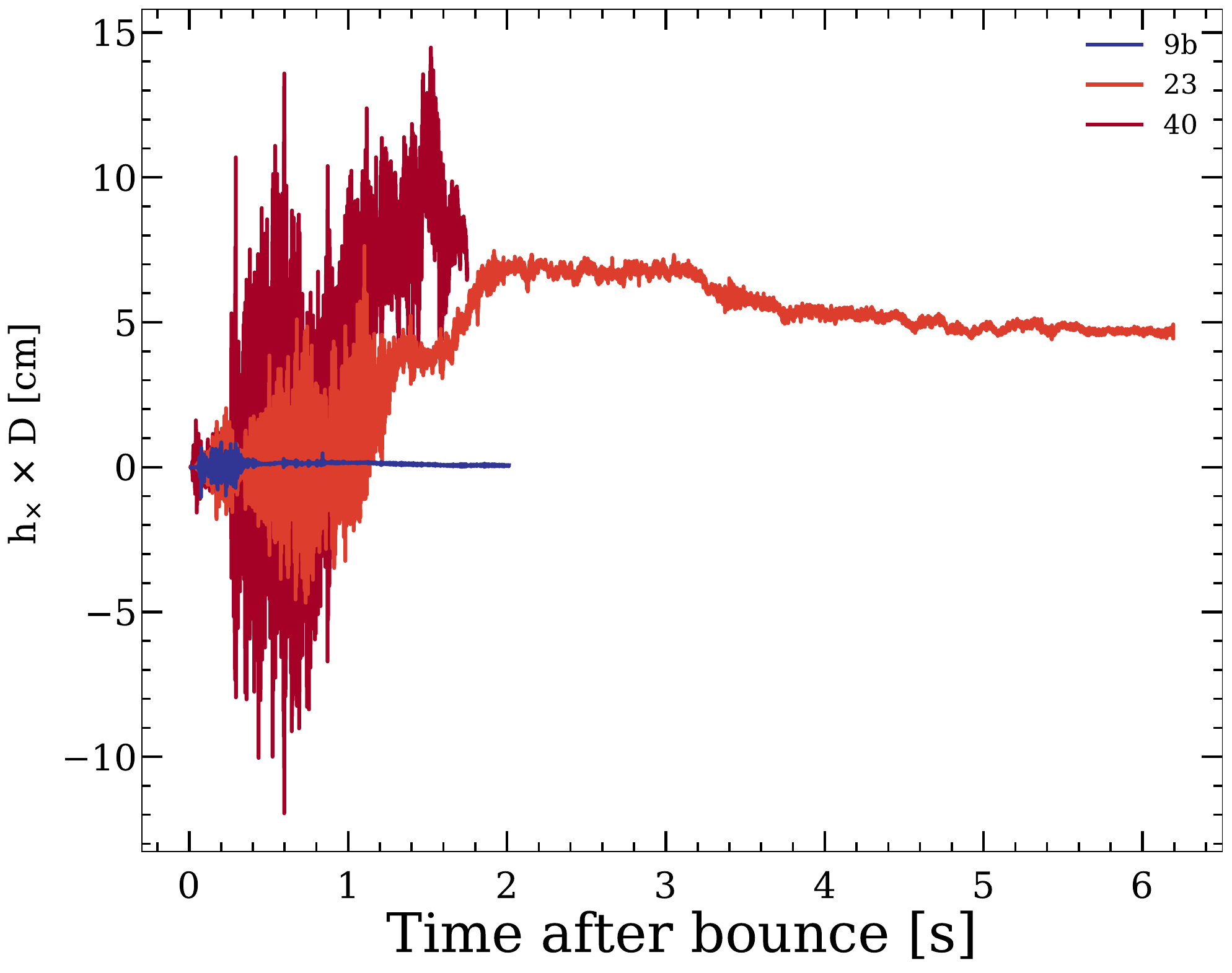}
    \caption{Gravitational wave strain for both $+$ (top) and $\times$ (bottom) polarizations as a function of time after bounce (in seconds) for two representative 3D models (the 9b- and 23-M$_{\odot}$), in comparison with the corresponding strains for the 40-M$_{\odot}$ model. This reflects a hierarchy in initial model compactness. Note the greater magnitude of the strain and memory (offset from zero) for the latter.}
    \label{fig:gw}
\end{figure}

\begin{figure}
    \centering
    \includegraphics[width=0.47\textwidth]{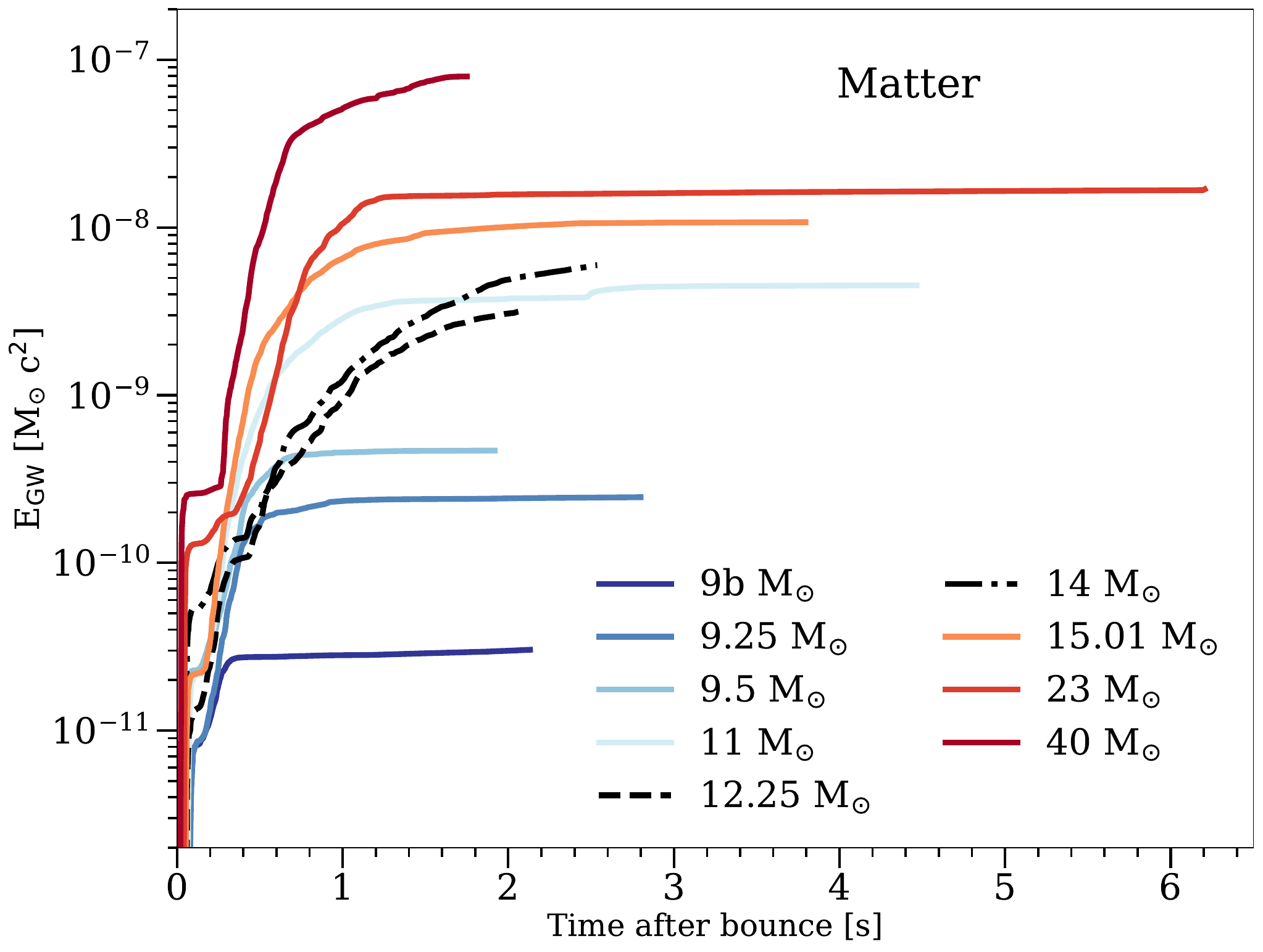}
        \includegraphics[width=0.47\textwidth]{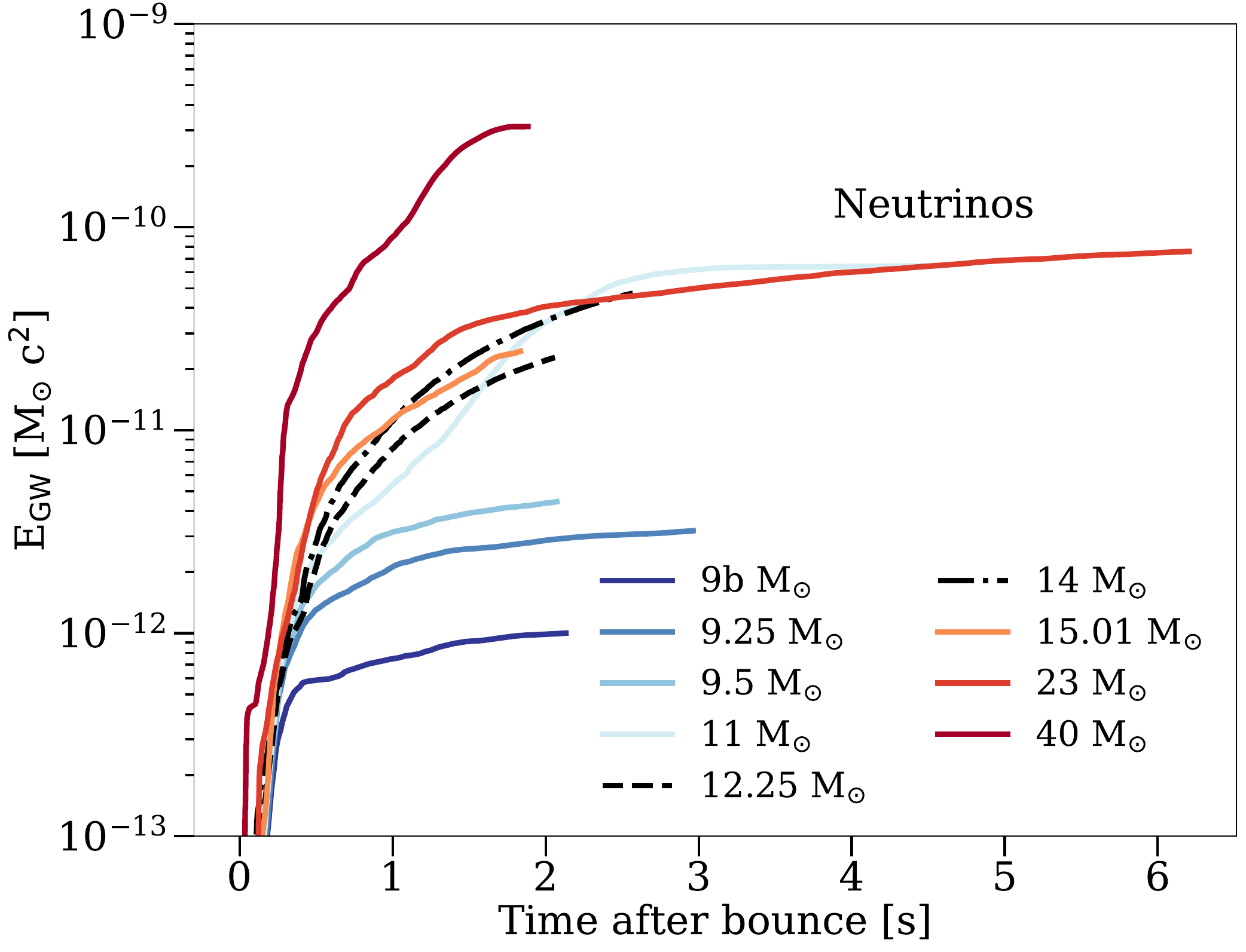}
    \caption{Gravitational wave energy radiated (in M$_{\odot}$c$^{2}$) from matter (left) and neutrinos (right) as a function of time after bounce (in seconds) for many comparison 3D models vis \`a vis the 40-M$_{\odot}$ model.}
    \label{fig:Egw_matter}
\end{figure}

\begin{figure*}
    \centering
    \includegraphics[width=0.47\textwidth]{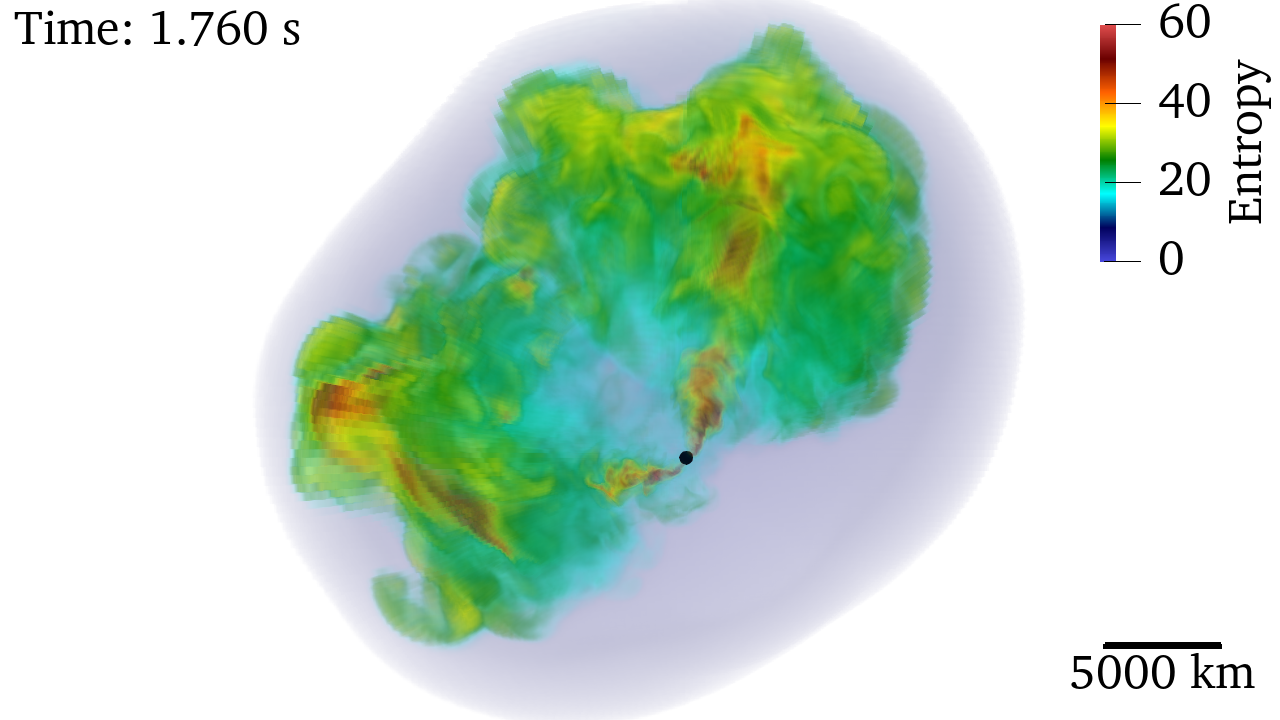}
    \includegraphics[width=0.47\textwidth]{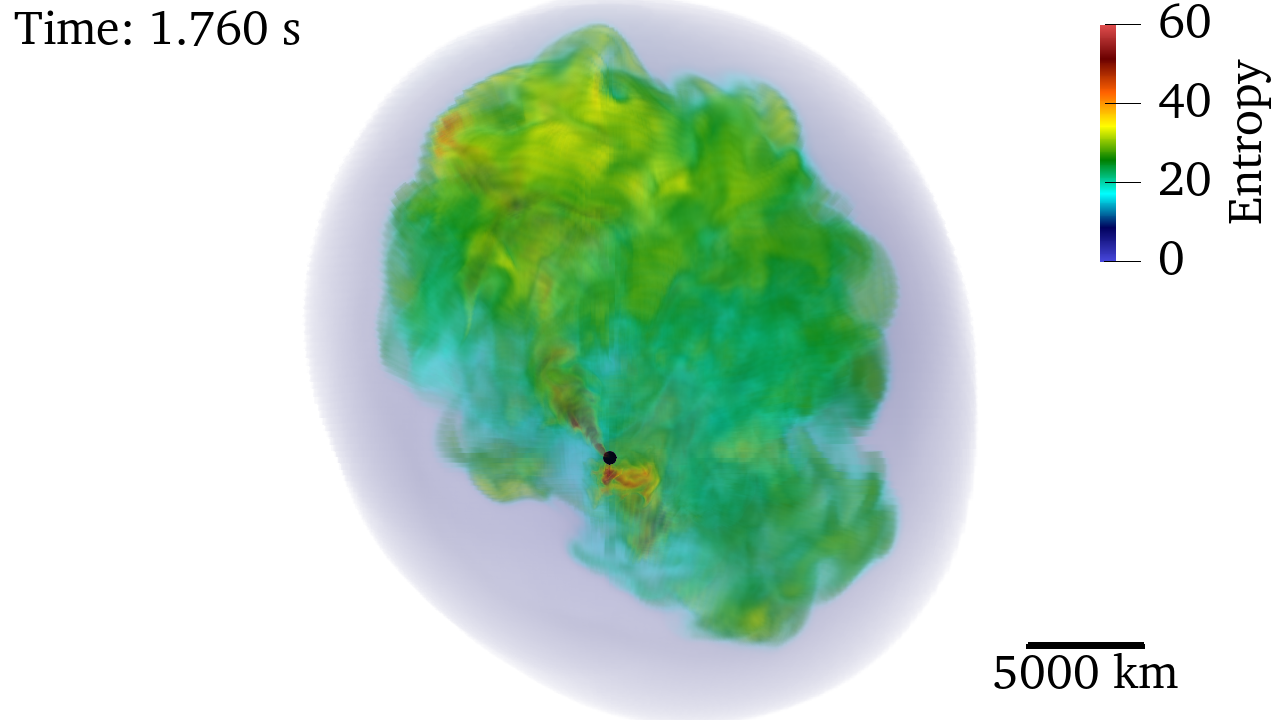}
    \caption{3D volume rendering of the explosion at 1.76 seconds after bounce and on a large scale. At this time the shock wave has achieved a distance of $\sim$20,000 km. The viewing angles in the two panels are perpendicular to each other. Color shows the entropy and the black dot shows the position of the PNS that will very soon transition to a black hole. The explosion is very asymmetrical, and two high-entropy jets originating from the central compact object are clearly seen (in red).}
    \label{fig:S-large-scale}
\end{figure*}

\begin{figure*}
    \centering
    \includegraphics[width=0.47\textwidth]{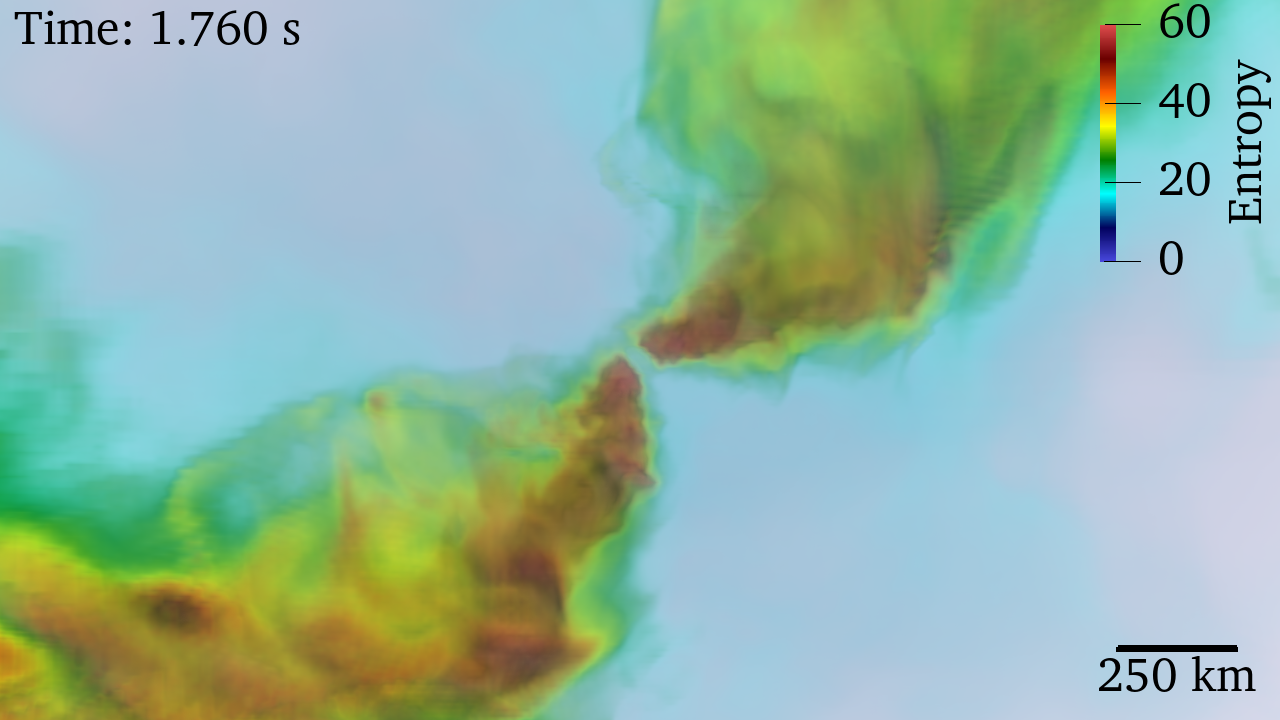}
    \includegraphics[width=0.47\textwidth]{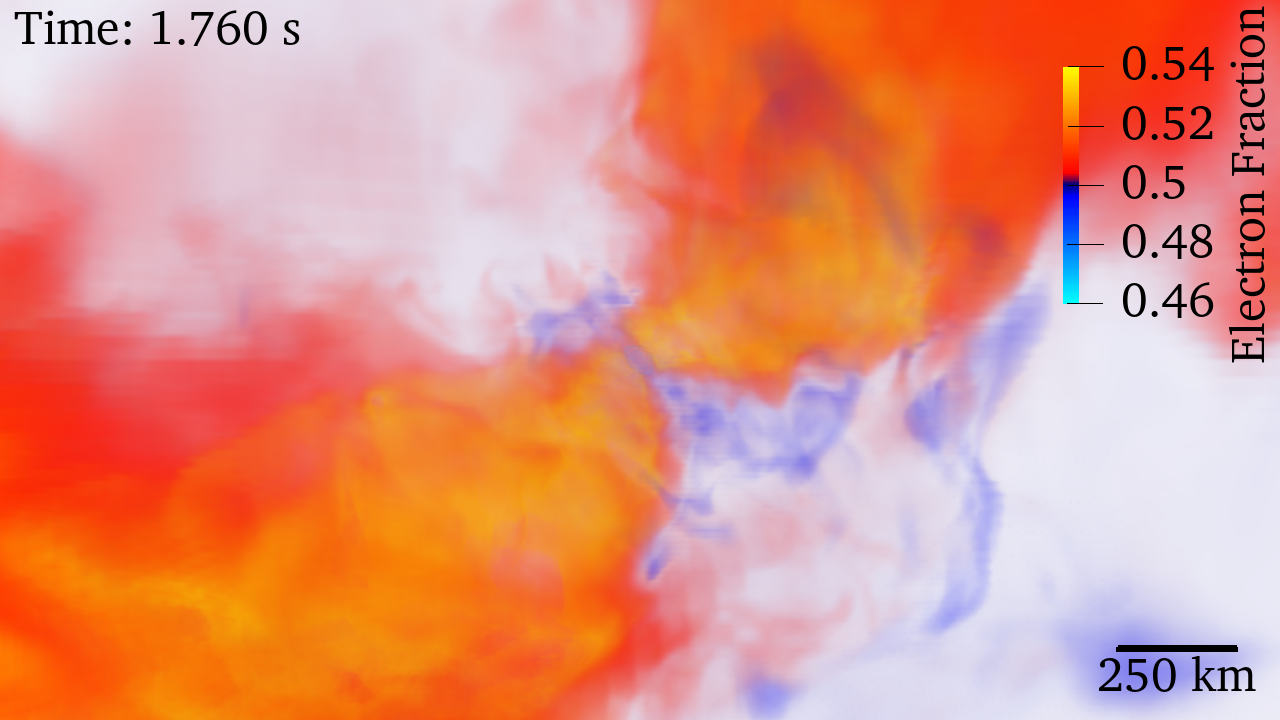}
    \includegraphics[width=0.47\textwidth]{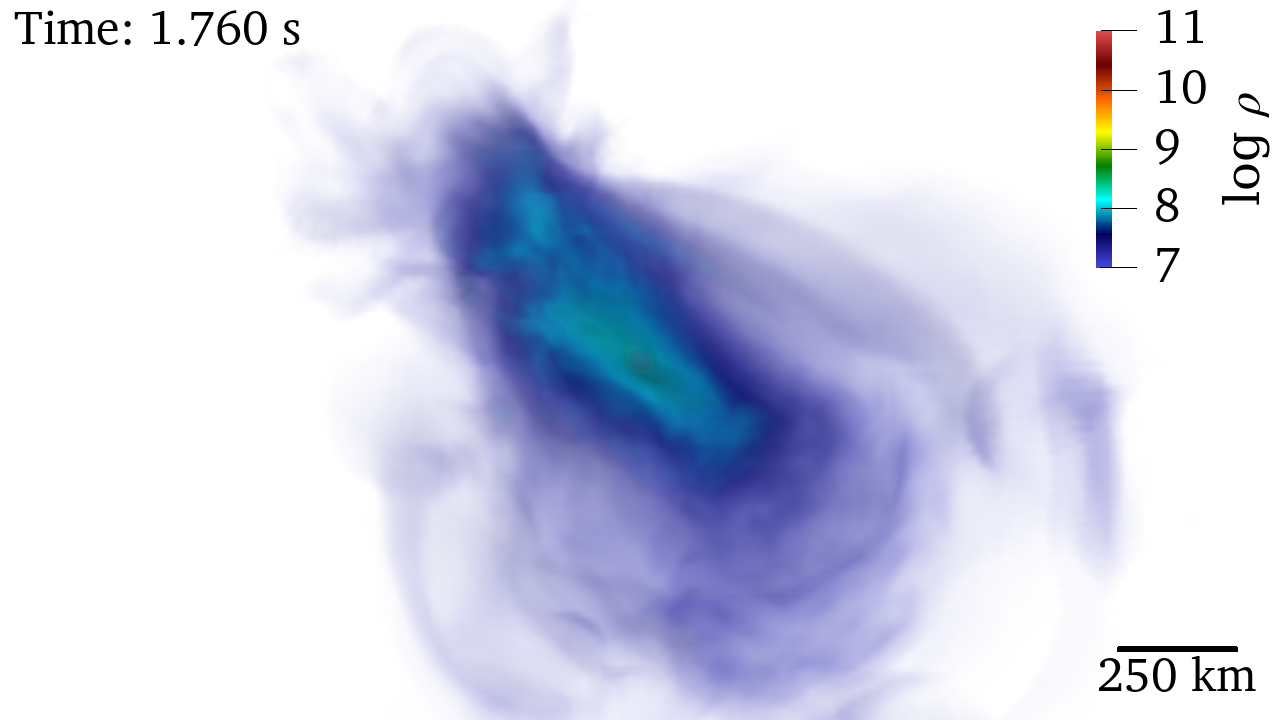}
    \includegraphics[width=0.47\textwidth]{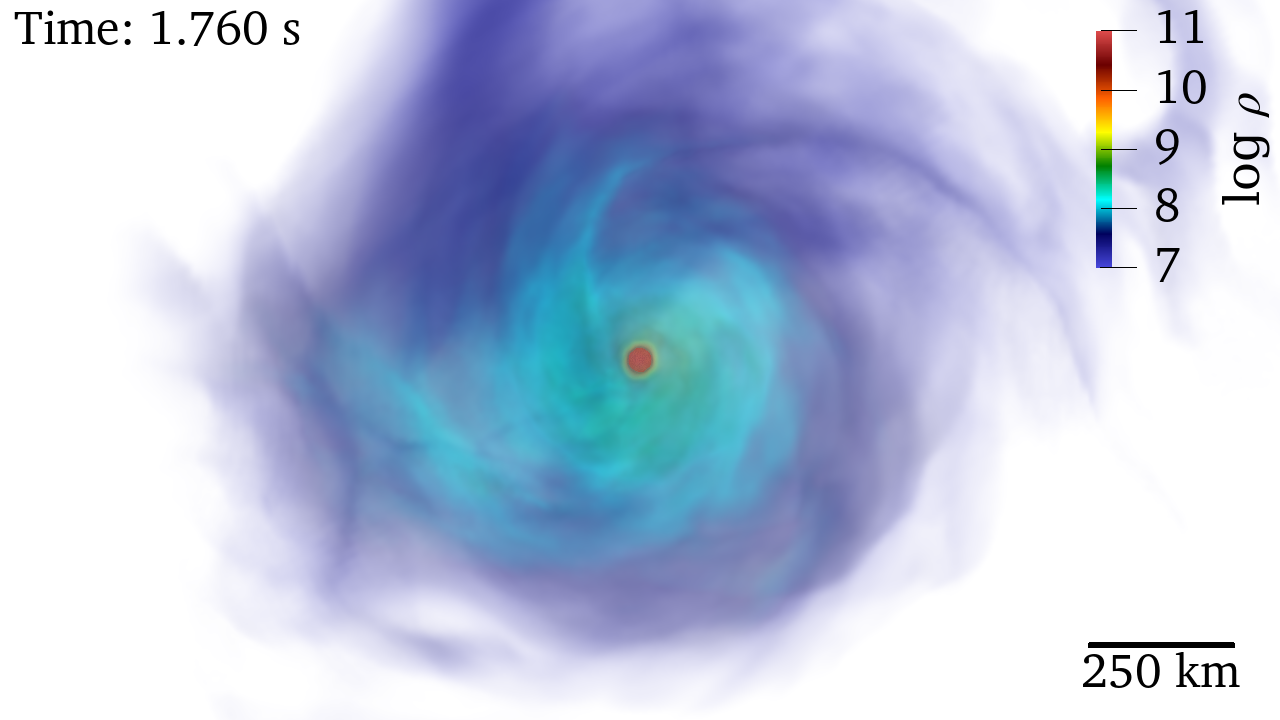}

    \caption{3D volume rendering of the explosion at 1.76 seconds after bounce showing the entropy, $Y_e$, and density of the central region. The viewing directions in all panels except the bottom right one are the same as in Figure \ref{fig:S-large-scale}, which is perpendicular to the general direction of the neutrino-driven jets. The viewing direction of the bottom right panel is along the jets. {\bf Top Left:} Entropy distribution in the central region. At this time, the high-entropy material moves out only along the jets. {\bf Top Right:} Electron fraction distribution in the central region. A jet and disk structure can be seen, in which the jets are proton-rich ($Y_e>0.5$) (orange red) and the disk is neutron-rich ($Y_e<0.5$) (blue). For this rendering, the $Y_e=0.5$ region is made more translucent. {\bf Bottom Left:} Density distribution in the central region with the same orientation as above. Only matter with $\rho>10^7$ g cm$^{-3}$ is shown here. The higher-density matter roughly traces the shape of the disk. {\bf Bottom Right:} Same as the bottom left panel, but as viewed down the axis of the top-right jet.}
    \label{fig:3d-edge-on}
\end{figure*}

As stated in \S\ref{rotation}, near the end of our full-radiation transport simulation a disk of matter, fed mass and angular momentum via the accretion of infalling plumes, enshrouds the inner core. Figure \ref{fig:S-large-scale} recapitulates in part Figure \ref{fig:ent}, but is volume-rendered and for two relatively perpendicular observation directions. As this figure suggests, there are two roughly opposite-pointing jets emerging from an inner region of relatively low entropy.
This inner region is seen much more clearly in the panels of Figure \ref{fig:3d-edge-on}. Note the scale bar. The top left panel depicts the entropy and clearly reveals jet structures with high entropy (red), surrounded by a lower-entropy annular region (light blue). The width of the base of the high-entropy jet is $\sim$200 km. Not seen here, but apparent in movies we have made, is the slight wobble of the thick disk. The top-right panel of Figure \ref{fig:3d-edge-on} depicts the electron fraction (Y$_e$) in the same region. Most interesting is the contrast between the lower-Y$_e$ matter (blue) of the disk and the higher-Y$_e$ matter (red) of the jets. The proton-richness of the jet is a natural consequence of net $\nu_e$ absorption (versus $\bar{\nu}_e$ absorption) that pushes Y$_e$ up from the low values with which it is launched. This is a well-understood phenomenon \citep{pruet2006,burrows_2020}. The lower Y$_e$ of the disk is a consequence of the fact that the matter is maintained at higher densities by staying behind in a circulating disk. It is also fed mass that has electron-captured on infall. This Y$_e$ contrast between disk and jet seems to be a robust feature. Two views of the mass density distribution of the disk itself are given in the lower left and right panels. The latter in particular makes clear the disk-like morphology that surrounds the inner core (seen here in red).

\begin{figure*}
    \centering
    \includegraphics[width=0.47\textwidth]{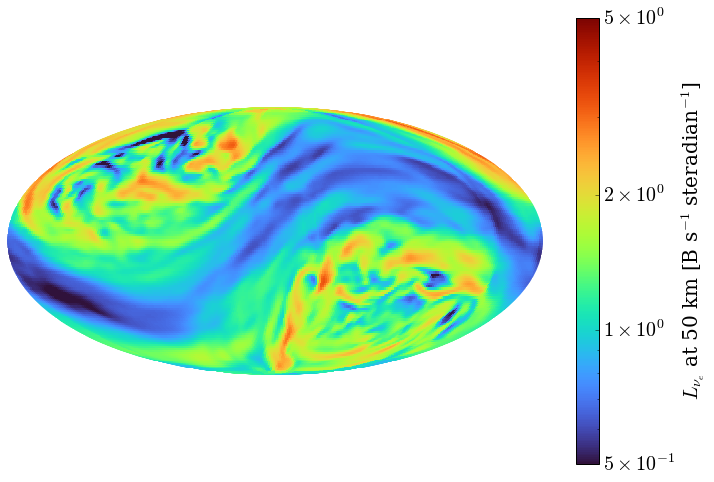}
    \includegraphics[width=0.47\textwidth]{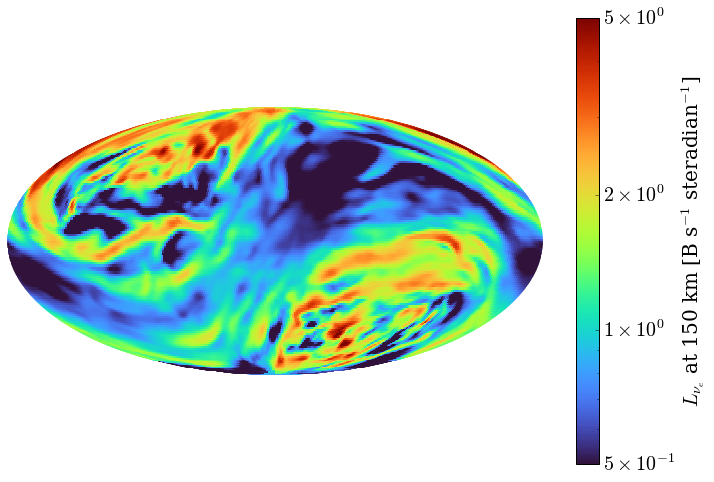}
    \caption{Luminosity per steradian of the electron-type neutrinos at 1.76 seconds after bounce. The two panels are measured at 50 and 150 km, respectively. In both panels we see that the enhanced opacity of the disk and the rotation of the outer PNS together lead to roughly dipolar neutrino emission.}
    \label{fig:Lnue-mollweide}
\end{figure*}

One of the most interesting features of this simulation is the anisotropy of the emergent neutrino emissions. The disk
partially obscures the equatorial regions. In so doing, the apparent radii of the neutrinospheres move out to lower temperatures in the equatorial band.  There is also a slight ``von Zeipel" effect of a rotating star \citep{zeipel}. As a consequence, most of the neutrino emission emerges from the poles. Figure \ref{fig:Lnue-mollweide} is a Molleweide projection at two different distances from the core of the electron-neutrino luminosity per steradian and demonstrates this result clearly. The pole-equator constrast is most pronounced and is the reason neutrino-driven jets arise in the first place during the later evolutionary phases. Again, we note that some of the jet matter reaches speeds of $\sim$45,000 km s$^{-1}$. This high-speed matter interacts with the earlier ejecta, generating secondary shocks. These secondary shocks are seen in Figures \ref{fig:ent} and \ref{fig:S-large-scale}. This phenomenon is similar to the wind/primary-ejecta interaction seen by \citet{wang_wind}. However, here the neutrino-driven wind is narrowly driven into two countervaling jets. 

\section{Post Black Hole Formation Evolution}
\label{post}

\begin{figure*}
    \centering
    \includegraphics[width=0.49\textwidth]{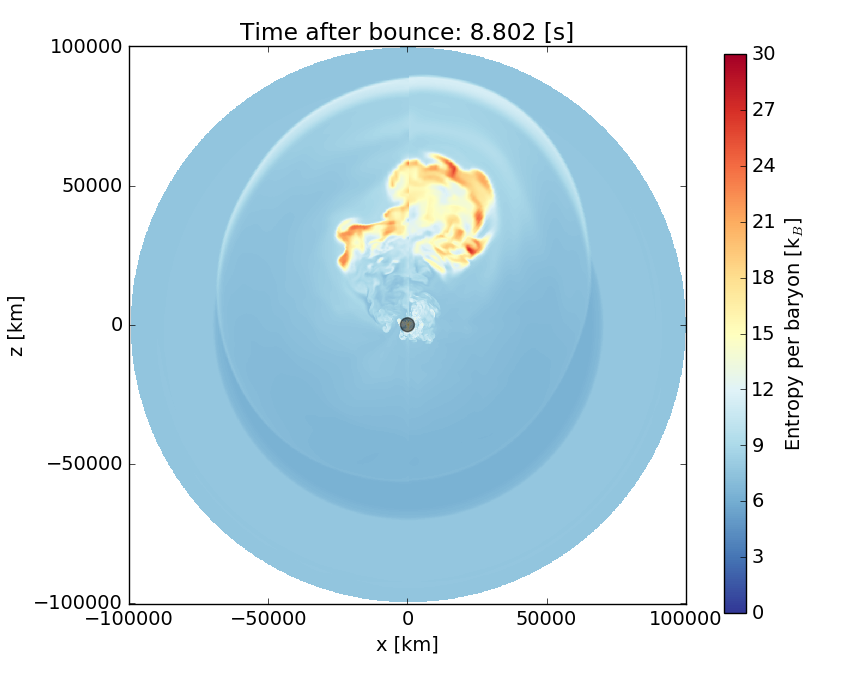}
    \includegraphics[width=0.49\textwidth]{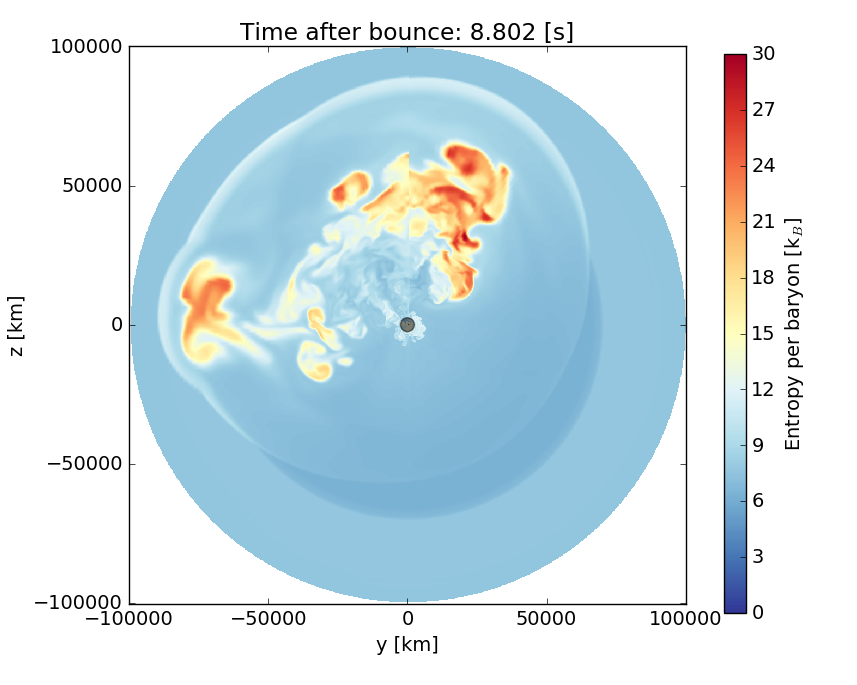}
    \includegraphics[width=1.0\textwidth]{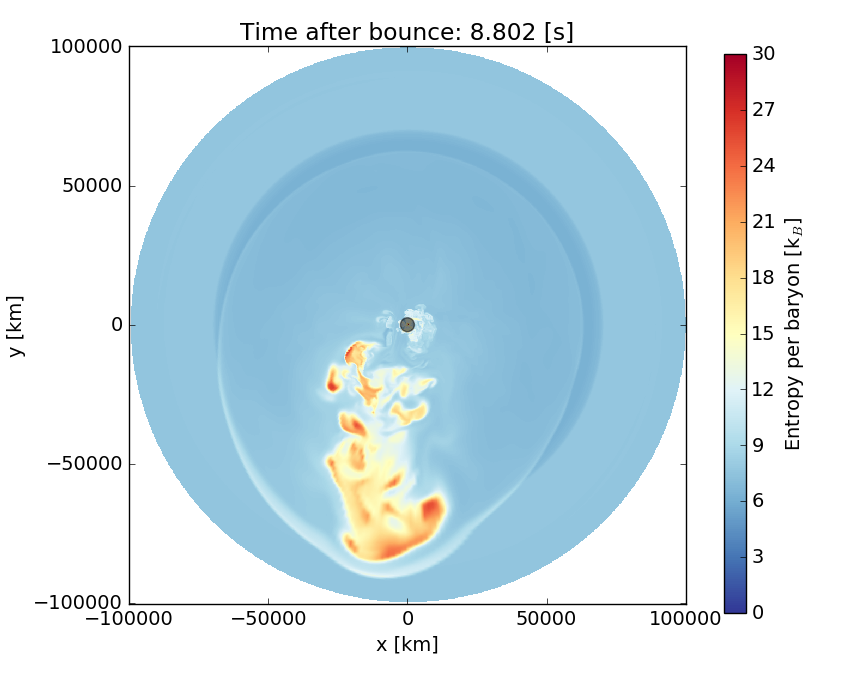}
    \caption{Entropy slices at $\sim$8.8 seconds post-bounce for the 40-M$_{\odot}$ solar-metallicity progenitor in the z-x, z-y, and y-x planes. The black dot marks the position of the black hole. Neutrino transport is turned off at 1.76 seconds after bounce, and a flow-in diode boundary condition is applied at 100 km. Our outer grid boundary is at 100,000 km. On this plot at this time the shock wave has penetrated the carbon/oxygen interface (darker blue to lighter blue transition near $\sim$60,000 km) in one sector (bottom of the y-x panel), while still interior to it in the other.}
    \label{fig:ent-late}
\end{figure*}

After the general-relativistic instability ensues and a black hole would form (within less than one millisecond), our code can not follow its subsequent evolution. However, at that point, we excise the core, replace it with a point mass, suppress neutrino transport, install a flow-in diode boundary condition at 100 kilometers, and then continue
the 3D simulation in F{\sc{ornax}} with only hydrodynamics and gravity. \adam{The diode boundary condition is akin to the standard flow-in boundary that does not allow matter to emerge, in our case from the 100 km inner boundary, during the post black-hole formation simulation phase.} We ensure that the lapse function before and after the mapping is continuous, so that the accelerations are continuous. Matter \adam{and angular momentum} that flows through the inner boundary is added to the point mass. The total number of grid points in 3D is $\sim$33,000,000, similar to the number of grid points for the preceding full-physics run.  In this way, we can and do follow the post-black-hole-formation evolution.  The goal of this post-simulation is to determine the
subsequent blast evolution and black hole growth to its asymptotic state.  We find that after about seven more seconds (to $\sim$8.8 seconds after bounce) the black hole has indeed accreted an additional $\sim$0.2 M$_{\odot}$ and achieved a baryon mass of $\sim$2.63 M$_{\odot}$ and a gravitational mass near 2.49 M$_{\odot}$.
Moreover, the additional accreted angular momentum in this mass boosts the relativistic spin parameter up to $\sim$0.19. That angular momentum is approximately along the axis of the stronger jet. The explosion energy has nearly plateaued at $\sim$1.6 Bethes, the difference with the previously-quoted value of $\sim$1.416 Bethes being due to the shedding of the negative contributions due to the matter that subsequently accreted into the black hole.  The formation of the black hole effectively cuts off neutrino heating, but the explosion is already well along. Hence, we find with this post-secondary collapse simulation that at $\sim$8.8 seconds post
bounce the shock wave has achieved a maximum radius of $\sim$90,000 km and shows no sign of abating. In particular, no rarefaction wave due to black hole formation alters the blast's expansion, except for the small amount of matter already tagged as moving in at the time of black hole formation (i.e., the ``shadow" at ``5 o'clock," see Figure \ref{fig:ent}). The composite of the pre- and post-black-hole formation phases provides a unique insight into this exotic black-hole formation/explosion channel. Figure \ref{fig:ent-late} depicts a 2D slice at $\sim$8.8 seconds after bounce of the 3D entropy field at this time.  One should compare this plot at $\sim$8.8 seconds with Figure \ref{fig:ent} at $\sim$1.75 seconds.

\adam{However, even $\sim$8.8 seconds after bounce, the newly-formed black hole is still accreting matter (mostly from the ``shadow" direction) at a rate of $\sim$2$\times$10$^{-2}$ M$_{\odot}$ s$^{-1}$ . Therefore, determining the final black hole mass will require simulations out to much later times. Though an explosion seems assured, the final black hole mass left behind could still be a large fraction of the progenitor star.}

\section{Conclusions}
\label{conclusions}

In this paper, we have simulated the collapse and subsequent evolution of the core of the solar-metallicity 40-M$_{\odot}$ progenitor model of \citet{swbj16} and found that it explodes vigorously by the turbulence-aided neutrino mechanism. This despite the fact that this model boasts the highest compactness of any star in the associated solar-metallicity massive-star model suite. Moreover, within $\sim$1.5 seconds of explosion, a black hole forms. This is a consequence of simultaneous ongoing accretion and explosion that is a generic feature of 3D CCSN explosion theory. The explosion is very asymmetrical, and at the time of black hole formation mass accretion at an instantaneous rate of $\sim$2$\times$10$^{-2}$ M$_{\odot}$ s$^{-1}$, mostly from one large sector, is seen. This is unlike the outcome often witnessed in recent 3D simulations for very-low- and zero-metallicity progenitors in this ZAMS mass range for which any explosion seems quickly followed by black hole formation, aborting the blast and stalling the shock wave into accretion. For such even-higher compactness models, the short time delay between explosion and black-hole formation is a consequence of the even larger post-bounce PNS mass growth rate. The result is that there is no time before the general-relativistic instability ensues for neutrino heating in the gain region to deposit enough energy to overcome the binding energy of the mantle overburden and the rest of the star. A large mass black hole, with a value perhaps set by the total stellar mass upon collapse, should then result.

Rather, for this solar-metallicity 40-M$_{\odot}$ ZAMS star, we find a total explosion energy of $\sim$1.6$\times$10$^{51}$ ergs. We also find an ejected $^{56}$Ni mass of $\sim$0.16 M$_{\odot}$, using the approach in \citet{wang_wind}. At the time of black-hole formation, the PNS baryon mass is $\sim$2.434 M$_{\odot}$ and the gravitational mass is 2.286 M$_{\odot}$.  After $\sim$7 seconds of further evolution, the black hole's baryon mass is $\sim$2.63 M$_{\odot}$ and gravitational mass is $\sim$2.48 M$_{\odot}$, but at $\sim$8.8 seconds the mass accretion rate is still $\sim$2$\times$10$^{-2}$ M$_{\odot}$ s$^{-1}$. The inferred gravitational mass at $\sim$8.8 seconds is the lowest mass black hole ever theoretically found using state-of-the-art 3D computational tools applied to massive-star progenitor models. We note, however, there is currently no observational support for such a low-mass black hole \citep{2023arXiv230409368M} \adam{and that the simulation needs to be conducted to much later times to determine whether mass accretion abates or continues, leaving open the possibility that a much more massive black hole might remain. This is an exciting goal for future research.  However, if our conclusion that a low-mass black hole in the mass range $\sim$2.5-3.0 M$_{\odot}$ forms in this context were to survive, black holes with such a provenance will in part fill the putative mass gap between observed neutron stars and black holes \citep{2023arXiv230409368M} that is a subject of current vigorous debate \citep{bailyn1998,ozel2010,farr2011,shao2022,mandel2021}.  We emphasize that the large associated kick we find would make their direct detection a challenge.} 

Moreover, we find that a disk forms around the PNS, from which a pair of neutrino-driven jets emanates. These jets accelerate some of the matter up to speeds of $\sim$45,000 km s$^{-1}$ and contain matter with entropies approaching $\sim$50 per baryon per Boltzmann's constant. Such high-speed matter decelerates at secondary shocks produced when the jets impinge upon the primary ejecta. This interaction is reminiscent of that seen generically in the production and interaction of PNS winds \citep{wang_wind}. The disk opacity and rotation lead to emergent neutrino luminosities before black-hole formation that are strongly bipolar, not surprisingly in the direction of the jets they drive. Furthermore, the large spatial asymmetry in the explosion results in a residual black hole recoil speed of $\sim$1150 km s$^{-1}$. This high speed will likely unbind the newly-birthed black hole from any binary in which its progenitor might have been found. 

\begin{figure*}
    \centering
    \includegraphics[width=0.47\textwidth]{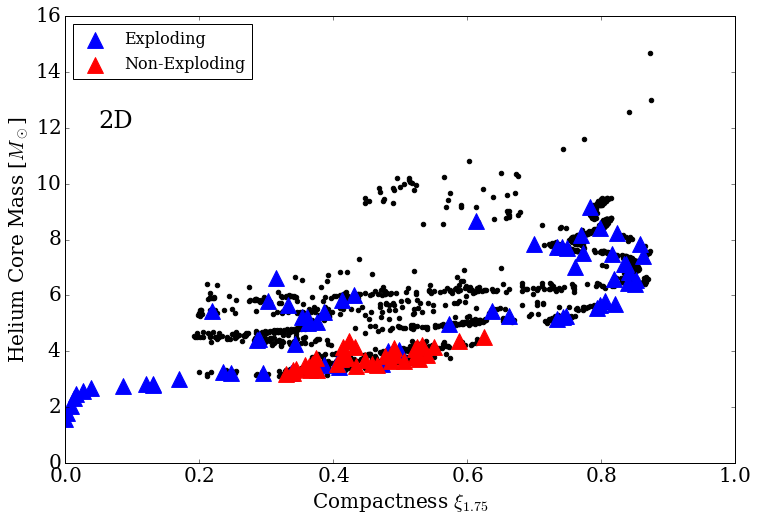}
    \includegraphics[width=0.47\textwidth]{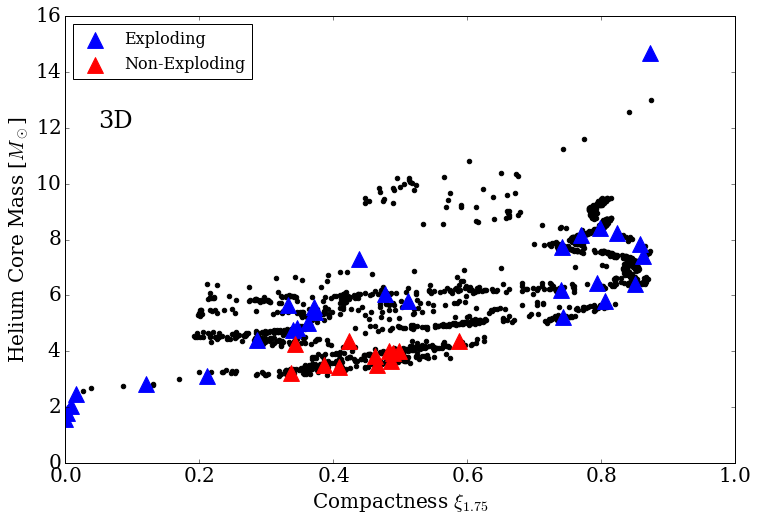}
    \caption{A mapping of the explodability of massive-star progenitors in compactness/helium-core-mass space. The red triangles are non-exploding models, the blue triangles are exploding models, and the black dots are 
    the \citet{swbj16} and \citet{sukhbold2018} solar-metallicity progenitor suite. The left panel is using the 2D simulations in \citet{wang} and \citet{tsang2022} and the right panel is for a representative set of the 3D F{\sc{ornax}} simulations we have performed since 2019.  Notice the island of non-explosion in this space, for both 2D and 3D simulations, suggesting a special black-hole forming branch. The 40-M$_{\odot}$ model highlighted in this paper is in the top right corner of each panel. We have yet to simulate this model in 2D.}
    \label{fig:compactness-he}
\end{figure*}

This novel black-hole formation channel now joins the other putative black-hole formation channel discovered via our previous 3D \citep{burrows_2019,burrows_2020} and 2D simulations \citep{wang,tsang2022,vartanyan2023b}. The two panels in Figure \ref{fig:compactness-he} of the helium core mass versus compactness summarize this black-hole formation channel. The black dots indicate the positions in this space of the \citet{swbj16} and \citet{sukhbold2018} solar-metallicity models. The red triangles denote models that did not explode, and presumably form black holes, while the blue triangles denote models that exploded. Note the various clear branches seen in the Figure \ref{fig:compactness-he}, in part a reflection of the complicated interaction of burning shells, of shell mergers, and of the transition from core to shell carbon burning at the terminal stages of massive star evolution \citep{swbj16}. Given the limitations of spherical stellar evolution models to capture such systematics, the positions of the black dots on Figure \ref{fig:compactness-he} should be considered quite provisional. They are nevertheless suggestive.

The important aspect of Figure \ref{fig:compactness-he} is that the duds are found to cluster on the same branch, and in both 2D and 3D. This is a branch for which the compactness is intermediate and the jump in density at the silicon/oxygen interface is muted. Between a third and two-thirds of the models on this branch do not explode. \adam{For all these non-exploding models, compact structures interior to a stalled shock, which has receded to small radius, result. Such structures have strong neutrino cooling components interior to the gain region and large pre-shock ram pressures which both help to undermine explosion. Moreover, in the roughly 200 2D and 3D simulations we to date have performed we have witnessed the emergence of a supernova after $\sim$1.2 seconds after bounce for only one model, which tettered for this long. Finally, we note that we have carried the 14 M$_{\odot}$ model shown in this paper out further to $\sim$2.4 seconds and it remains a dud.}.  However, it should be stressed that this conclusion is contingent upon the fact that the associated simulations are all done by us using our F{\sc{ornax}} code. Future work could well improve, alter, or overthrow this conclusion. However, we suggest that at this stage in the development of CCSN theory this conclusion may reflect the best current thinking on the origin from solar-metallicity stars of stellar-mass black holes. To be sure, the actual residual mass of such black holes will depend on the various mass loss mechanisms in and out of tight binaries, and in the last stages leading to collapse, such as the pre-supernova mass loss inferred in recent supernova light curve studies \citep{2015ApJ...814...63M} and implied by the ``Nadezhin" effect \citep{nadezhin1980,lovegrove2013}. Nevertheless, the black-hole formation channel summarized in Figure \ref{fig:compactness-he} is intriguing.

At solar metallicity, to this channel has now been added a second with our simulation of the 40-M$_{\odot}$ model. As stated in \S\ref{basics}, at very-low- and ``zero-metallicity" in and above this mass range, black hole formation without a supernova explosion is the likely outcome, and the black hole masses that would emerge would be much larger. These black holes may be many of the black holes seen in the growing LIGO database \citep{aligo}, with some degree of subsequent hierarchical growth by merger also likely.

Nevertheless, we can estimate the fraction of massive stars more massive than $\sim$8 M$_{\odot}$ that might form black holes.  Using the Salpeter mass function, we see that $\sim$15\% of massive stars reside between 12- and 15-M$_{\odot}$.  If two-thirds of these stars form black holes, this constitutes $\sim$10\% of massive stars.  If we arbitrarily group the stars between 30 and 50 M$_{\odot}$ as residing with the 40-M$_{\odot}$ star in this second formation channel, this constitutes another $\sim$8$-$10\%.  The suggestion would be that approximately 20\% of solar-metallicity massive stars would leave black holes, but that the character of these black holes would be quite different. One channel would leave black holes in perhaps the $\sim$5-15 M$_{\odot}$ range, and with low kick speeds, while the other might leave black holes in $\sim$2.5-3.0 M$_{\odot}$ mass range \adam{(though this is still to be determined, see \S\ref{post})} with high kick speeds. The former would be associated with the black hole population seen in X-ray binaries and using Gaia \citep{2023arXiv230409368M}. The latter, we believe, is yet to have observational support. 

Importantly, it should be borne in mind that our results for the 40-M$_{\odot}$ are contingent in part upon the nuclear equation of state employed by us (here, SFHo) and the stellar mass-loss prescription employed by \citet{swbj16}. A different nuclear EOS would have resulted in an altered delay between explosion and black hole formation. Even if another nuclear EOS led to black hole formation, the delay between explosion and black hole formation would have been different. This would have  consequences for, among other things, the energy of explosion. Therefore, the scenario which we articulate in this paper depends, however weakly, on the nuclear EOS in perhaps interesting ways. For instance, had we employed the DD2 EOS, for which the maximum cold neutron-star mass is 2.93 M$_{\odot}$ (baryonic) and 2.42 M$_{\odot}$ (gravitational), though the high compactness would still likely have led to an explosion, a black hole may not have formed, since accretion after explosion would not have been sufficient to push the PNS core into instability. A massive neutron star would have been the product.  However, the DD2 EOS also predicts larger neutron star radii, which are mildly at odds with current observational constraints \citep{2013ApJ...774...17S,2013ApJ...765L...5S}.  

A different stellar mass-loss formulation, Roche-lobe overflow, or common-envelope evolution, would have altered the core's structure and compactness at primary collapse and the binding energy of the overburden. The case of a zero-metallicity, isolated progenitor was discussed in \S\ref{previous}. Therefore, the precise evolutionary path followed by the progenitor is equally germane. Reversing these caveats, both the EOS and the mass-loss dependencies offer means to constrain both, should the objects we suggest can be created in this scenario ever be detected.  

It is important to list some of the limitations of our computations.  First, as stated in \S\ref{method}, our code employs approximate general relativity, but a fully general-relativisitc code is called for that can also transition physically through black hole formation to continue the subsequent evolution seamlessly \citep{2023arXiv230706192K}. Second, we see the creation of a rapidly rotating disk, but have not addressed the possible effects of magnetic fields. We find that on longer timescales the angular momenta sufficient to power a generic collapsar may be available, and that there can be little doubt that some sort of MHD-driven jet should arise after black hole formation. \adam{This would imply that a collapsar-like gamma-ray burst could emerge from a massive star that is initially non-rotating or slowly-rotating, contrary to the suggested context of the original collapsar paradigm \citep{1993ApJ...405..273W}.} This should be an important topic of future study. In addition, we have not considered the effects of an initially rotating profile. The rotation we see is induced by the hydrodynamics and whether this is larger than, comparable to, or smaller than core rotation at stellar death is yet to be determined.  Finally, we have predicted that this channel of black hole formation leads to a supernova. The electromagnetic signature of this explosion and its morphology might be unique and these aspects need to be fully addressed.


\section*{Data Availability}

\adam{The data presented in this paper can be made available upon reasonable request to the first author.  However, we have also deposited in the \url{https://doi.org/10.7910/DVN/XAHJRS} repository the data need to generate Figure 7 and an associated movie to enable readers to explore these data independently.}

\section*{Acknowledgments}

We thank Christopher White, Matthew Coleman, and Benny Tsang for valuable discussions throughout the course of this project. DV acknowledges support from the NASA Hubble Fellowship Program grant HST-HF2-51520. AB acknowledges support from the U.~S.\ Department of Energy Office of Science and the Office of Advanced Scientific Computing Research via the Scientific Discovery through Advanced Computing (SciDAC4) program and Grant DE-SC0018297 (subaward 00009650) and support from the U.~S.\ National Science Foundation (NSF) under Grant AST-1714267. We are happy to acknowledge access to the Frontera cluster (under awards AST20020 and AST21003). This research is part of the Frontera computing project at the Texas Advanced Computing Center \citep{Stanzione2020}. Frontera is made possible by NSF award OAC-1818253. Additionally, a generous award of computer time was provided by the INCITE program, enabling this research to use resources of the Argonne Leadership Computing Facility, a DOE Office of Science User Facility supported under Contract DE-AC02-06CH11357. Finally, the authors acknowledge computational resources provided by the high-performance computer center at Princeton University, which is jointly supported by the Princeton Institute for Computational Science and Engineering (PICSciE) and the Princeton University Office of Information Technology, and our continuing allocation at the National Energy Research Scientific Computing Center (NERSC), which is supported by the Office of Science of the U.~S.\ Department of Energy under contract DE-AC03-76SF00098.





    \label{fig:profs}

%

\clearpage

\bibliographystyle{aasjournal}
\bibliography{References}

\label{lastpage}
\end{document}